\documentclass[aps,pra,twocolumn,10pt,amsmath,superscriptaddress,longbibliography,amssymb]{revtex4-1}
\usepackage{mathrsfs}
\usepackage{graphicx}
\usepackage{amsmath}
\usepackage{CJK}
\usepackage{amssymb}
\usepackage{amsfonts}
\usepackage{color}
\usepackage{multirow}

\usepackage{dcolumn}
\usepackage{bm}
\usepackage[dvipdfm, pdfstartview=FitH, CJKbookmarks=true, bookmarksnumbered=true, bookmarksopen=true, colorlinks=true, pdfborder=001, citecolor=blue, linkcolor=blue, linktocpage=true] {hyperref}

\begin{document}

\title{Compact gravimeter with an ensemble of  ultracold atoms in spin-dependent optical lattices}

\author{Yongguan Ke}
\affiliation{Laboratory of Quantum Engineering and Quantum Metrology, School of Physics and Astronomy, Sun Yat-Sen University (Zhuhai Campus), Zhuhai 519082, China}
\affiliation{State Key Laboratory of Optoelectronic Materials and Technologies, Sun Yat-Sen University (Guangzhou Campus), Guangzhou 510275, China}

\author{Jiahao Huang}
\affiliation{Laboratory of Quantum Engineering and Quantum Metrology, School of Physics and Astronomy, Sun Yat-Sen University (Zhuhai Campus), Zhuhai 519082, China}

\author{Min Zhuang}
\affiliation{Laboratory of Quantum Engineering and Quantum Metrology, School of Physics and Astronomy, Sun Yat-Sen University (Zhuhai Campus), Zhuhai 519082, China}
\affiliation{State Key Laboratory of Optoelectronic Materials and Technologies, Sun Yat-Sen University (Guangzhou Campus), Guangzhou 510275, China}

\author{Bo Lu}
\altaffiliation{Email: lubo3@mail.sysu.edu.cn}
\affiliation{Laboratory of Quantum Engineering and Quantum Metrology, School of Physics and Astronomy, Sun Yat-Sen University (Zhuhai Campus), Zhuhai 519082, China}

\author{Chaohong Lee}
\altaffiliation{Email: lichaoh2@mail.sysu.edu.cn}
\affiliation{Laboratory of Quantum Engineering and Quantum Metrology, School of Physics and Astronomy, Sun Yat-Sen University (Zhuhai Campus), Zhuhai 519082, China}
\affiliation{State Key Laboratory of Optoelectronic Materials and Technologies, Sun Yat-Sen University (Guangzhou Campus), Guangzhou 510275, China}
\affiliation{Synergetic Innovation Center for Quantum Effects and Applications, Hunan Normal University, Changsha 410081, China}

\date{\today}

\begin{abstract}
  Atomic interferometry in optical lattices is a new trend of developing practical quantum gravimeter.
  Here, we propose a compact and portable gravimetry scheme with an ensemble of ultracold atoms in gravitationally tilted spin-dependent optical lattices.
  The fast, coherent separation and recombination of atoms can be realized via polarization-synthesized optical lattices.
  The input atomic wavepacket is coherently split into two parts by a spin-dependent shift and a subsequent $\frac{\pi}{2}$ pulse.
  Then the two parts are held for accumulating a relative phase related to the gravity.
  Lastly the two parts are recombined for interference by a $\frac{\pi}{2}$ pulse and a subsequent spin-dependent shift.
  The $\frac{\pi}{2}$ pulses not only preclude the spin-dependent energies in the accumulated phase, but also avoid the error sources such as dislocation of optical lattices in the holding process.
  In addition, we develop an analytical method for the sensitivity in multi-path interferometry.

\end{abstract}

\maketitle

\section{Introduction}~\label{Sec1}
Precision gravity measurement is of great significance in geophysical applications, metrology and fundamental physics~\cite{Angelis2009}.
In addition to various classical gravimetry schemes~\cite{Richter1995,Niebauer1995,John1999,Jiang2012}, such as spring gravimeter and free falling corner cube gravimeter, quantum interferometry can be used to implement gravimetry.
The precision quantum gravimetry has been demonstrated via atom interferometry in free space, but it requires long free-fall distance and thus lacks mobility~\cite{Kasevich1992,Peters1999,Hu2013,Dickerson2013}.
Developing a compact and portable quantum gravimeter will have more practical applications~\cite{Ferrari2006,Hughes2009,Beaufils2011,Cadoret2012,Andia2013,Hamilton2015,Abend2016}.
A natural idea is trapping the atoms in optical lattices along the direction of gravity.
Then, the magnitude of the gravitational field can be determined via probing Bloch oscillations~\cite{Wannier1960,Ferrari2006}, using driven resonant tunneling of cold atoms~\cite{Ferrari2006,Poli2011,Tarallo2012}, or Wannier-Stark interferometry~\cite{Beaufils2011,Pelle2013}.
Atomic interferometry in optical lattices is becoming a new trend of developing a compact and portable quantum gravimeter.

A key process of atomic interferometry in optical lattices is to coherently split and recombine atoms in different heights.
In the Wannier-Stark interferometry, the beam splitter is realized by two-photon Raman transition~\cite{Beaufils2011,Pelle2013}.
To separate the atoms in large distance, one needs to increase the efficiency of Raman transition by decreasing the depth of optical lattices.
However, the decreasing of depth also decreases the number of trapped atoms.
There is a trade-off between distance separation, the interferometry time and the atomic number, which hinders the precision improvement.
The beam splitter can be alternately realized with spin-dependent optical lattices~\cite{Mandel2003,Steffen2012}.
Recently, polarization-synthesized optical lattices enable the fast, coherent and spin-dependent transport of atoms with large distance and high precision~\cite{Robens2017,Robens2018}.
The polarization-synthesized optical lattices may be an improved beam splitter, where the large distance and large atom number can be achieved at the same time.
A question naturally arises: Can we realize compact quantum gravimeter in polarization-synthesized spin-dependent optical lattices?

In this paper, we propose a compact gravimetry scheme via an ensemble of ultracold atoms in gravitationally titled spin-dependent optical lattices.
The input atomic wavepacket is coherently split into two parts at different heights by the first beam splitter (BS), which is achieved by tuning polarizations of the spin-dependent optical-lattice potential and then applying a $\frac{\pi}{2}$ pulse.
Holding the optical lattices for a certain duration of time, the two parts will accumulate a relative phase related to the gravity.
Then the two parts are recombined by the second BS via the inverse process of the first BS.
At last, the gravitational acceleration is extracted by applying a $\frac{\pi}{2}$ pulse and then measuring the spin population.
Importantly, through introducing a $\frac{\pi}{2}$ pulse in each BS, we remove the error sources from the dislocation of optical-lattice potential in the holding process.
The $\frac{\pi}{2}$ pulses can also be used to improve the precision of the single-atom digital interferometer~\cite{Steffen2012}.

This article is constructed as follows.
In Sec.~\ref{Sec1}, we briefly introduce the background and our motivation.
In Sec.~\ref{Sec2}, we describe our system of cold atoms in titled spin-dependent optical lattices.
In Sec.~\ref{Sec3}, we present the single-particle gravimetry scheme via interferometry in optical lattices.
In Sec.~\ref{Sec4}, we present the multi-particle gravimetry scheme via interferometry of coherent spin states (CSS's) and  SSS's.
In Sec.~\ref{Sec6}, we summarize our results and discuss the experimental possibility and potential applications.

\begin{figure}[!htp]
\begin{center}
\includegraphics[width=0.8\columnwidth]{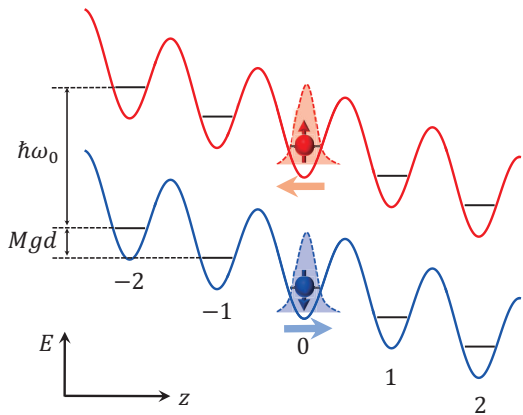}
\end{center}
  \caption{Schematic diagram of the spin-dependent Wannier-Stark (WS) systems. The spin-up and spin-down WS ladders parallel to each other with energy separation $\hbar\omega_0$ and equal spacings $Mgd$. Through adiabatically decreasing (increasing) the polarization phases $\varphi_{\uparrow (\downarrow)}$, the spin-up and spin-down atoms will be transported to higher and lower positions, respectively.}
\label{lattice}
\end{figure}

\section{Spin-dependent Wannier-Stark system}~\label{Sec2}

We consider an ensemble of two-level atoms $\{\left|\uparrow\right\rangle, \left|\downarrow\right\rangle\}$ within a deep spin-dependent optical-lattice potential aligned along the gravity direction.
The system obeys the Hamiltonian,
\begin{eqnarray}
\hat H =\left(
\begin{matrix}
\frac{{\hat p}_z^2}{2M}+ U_{\uparrow} + \frac{\hbar \omega_0}{2} & 0 \\
  0   & \frac{{\hat p}_z^2}{2M}+ U_{\downarrow}- \frac{\hbar \omega_0}{2}
\end{matrix}
\right)-Fz.
\end{eqnarray}
Here, $M$ is the atomic mass, $\hat p_z$ denotes the momentum along the gravitational direction, $\omega_0$ is the transition frequency between the two levels, and $U_{\uparrow(\downarrow)}$ is the spin-up (down) optical-lattice potential.
In the last term, $F=Mg$ is the gravitational force with the gravitational acceleration $g$.
Usually, the spin-dependent optical lattices~\cite{Mandel2003,Steffen2012,Robens2017} are described by $U_{\sigma}(z)=V_{\sigma}\sin^2(\kappa z-\varphi_{\sigma})$ with the amplitudes $V_{\sigma}$ and the phases $\varphi_{\sigma}$ for $\sigma=\{\uparrow,\downarrow\}$, where $\kappa=2\pi/\lambda$ is the common wave vector with the wavelength $\lambda$ (the lattice constant $d=\lambda/2$).

The single-particle eigenstates are known as the Wannier-Stark (WS) states~\cite{Wannier1960,GLUCK2002,Ferrari2006}, and their eigenvalues form the equidistant WS ladders
\begin{equation}
E_{\sigma,\alpha,l}=\epsilon_{\sigma,\alpha}-Fd( l+\varphi_{\sigma}/\pi)\pm \hbar \omega_0/2,
\end{equation}
with the band index $\alpha$ and the lattice site index $l$.
Here, $\epsilon_{\sigma,\alpha}$ is the bare on-site energy of the $\alpha$-th band in $\left|\sigma\right\rangle$.
We show the schematic diagram of the spin-dependent WS system in Fig.~\ref{lattice}.
The inner equidistant energy is $Mgd$ for both spin-up and spin-down WS ladders.
When the phases are adiabatically tuned according to $(\varphi_{\uparrow},\varphi_{\downarrow})=(-\nu t, +\nu t)$ with the driven frequency $\nu$,
the atoms in $\left|\sigma\right\rangle$ will follow the instantaneous WS energies $E_{\sigma,\alpha,l}(t)$.
Thus the spin-up and spin-down components will be shifted to higher and lower positions, respectively.
Furthermore, the atoms in a superposition of the two spin states will be coherently split into a superposition of two spatial wavepackets at different heights.
Next, we will show how to integrate the coherent spin-dependent transport into our interferometry scheme.

\begin{figure*}[!htp]
\begin{center}
\includegraphics[width=2\columnwidth]{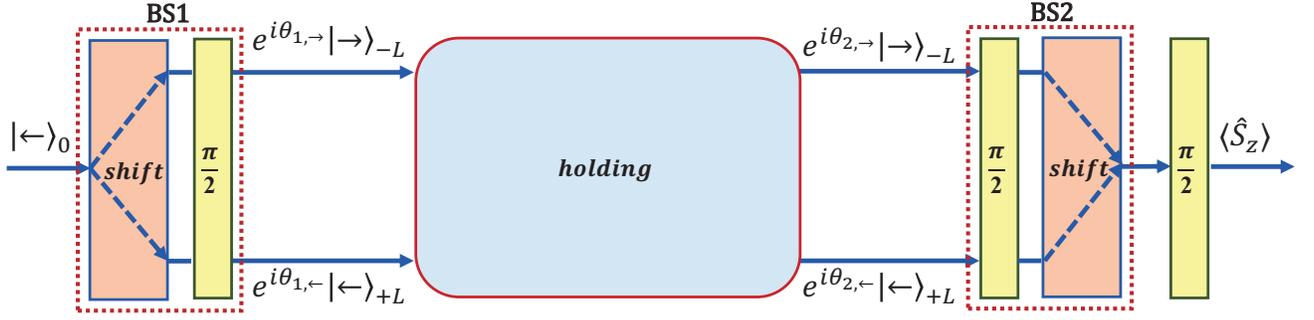}
\end{center}
  \caption{Gravimetry via Mach-Zehnder interferometry of a single atom in tilted optical lattices.
  The input state is prepared as  $\left|\leftarrow\right\rangle_{0}$ at the $0$-th lattice site.
  Through the first beam splitter (BS1), which contains a spin-dependent shift and a $\frac{\pi}{2}$ pulse, the input wavepacket is coherently split into two parts at $(-L)$-th and $(+L)$-th lattice sites (i.e. a superposition state of $\left|\leftarrow\right\rangle_{+L}$ and $\left|\rightarrow\right\rangle_{-L}$).
  In the holding process, the two parts at different heights will accumulate a relative phase due to the gravity, while the internal states rotate around the $\hat J_z$ axis due to the energy difference between $\left|\uparrow\right\rangle$ and $\left|\downarrow\right\rangle$.
  When the internal states of the two parts respectively rotate back to $\left|\rightarrow\right\rangle_{-L}$ and $\left|\leftarrow\right\rangle_{+L}$, we apply the second beam splitter (BS2 - the inverse process of BS1), the two parts are respectively flipped into $\left|\downarrow\right\rangle_{-L}$ and $\left|\uparrow\right\rangle_{+L}$, and then are shifted to the $0$-th lattice site for interference.
  At last, we apply a $\frac{\pi}{2}$ pulse and measure the spin population difference $\langle\hat J_z\rangle$.}
\label{systematic}
\end{figure*}

\section{Single-particle gravimetry scheme}~\label{Sec3}

Our gravimetry scheme is based upon a Mach-Zehnder interferometry, which includes a phase accumulation process (i.e. the holding process) sandwiched by two beam splitters (BS's), see Fig.~\ref{systematic}.
The first BS, which splits the input wavepacket into two parts at different heights, is achieved by a spin-dependent shift and a subsequent $\frac{\pi}{2}$ pulse.
In the holding process, the two parts accumulate a relative phase related to the gravity.
Then the second BS, which is the reverse process of the first BS, recombines the two parts for interference.
Lastly the relative phase is extracted by the spin population measurement after applying a $\frac{\pi}{2}$ pulse.
Obviously, our procedure is different from the digital atom interferometer~\cite{Steffen2012}, in which the shift operations of alternated direction are interleaved with $\pi$ pulses.
One may think the $\frac{\pi}{2}$ pulses in our BS's are not necessary, we will explain their importance later.

Given the creation (annihilation) operators $a^{\dag}_{\uparrow, l}$ ($a^{}_{\uparrow, l}$) and the Pauli matrices $\sigma_{x,y,z}$, our system can be described by the collective spin operators $\hat J_{x,y,z}=\frac{1}{2} \sum_{l} (a^{\dag}_{\uparrow, l}, a^{\dag}_{\downarrow, l}) \hat \sigma_{x,y,z} (a^{}_{\uparrow, l}, a^{}_{\downarrow, l})^{T}$.
Below, we will use the collective spin operator to analytically derive the accumulated phase, the spin population difference and its variance.

We first show how the phase is accumulated in the single-particle scheme.
Since only the $\{-l,0,+l\}$-th lattice sites are involved, the single-particle state can be written in the subspace as
\begin{equation}
\left|\psi\right\rangle= \left( {\begin{matrix}
{{w_{ \uparrow , - l}}} \\
{{w_{ \downarrow , - l}}}
\end{matrix}} \right)_{-l} \oplus \left( {\begin{matrix}
{{w_{ \uparrow ,0}}} \\
{{w_{ \downarrow ,0}}}
\end{matrix}} \right)_{0} \oplus \left( {\begin{matrix}
{{w_{ \uparrow ,+l}}} \\
{{w_{ \downarrow ,+l}}}
\end{matrix}} \right)_{+l},
\end{equation}
where $w_{ \sigma ,l}$ is amplitude of the Wannier-Stark state $\left|W_{\sigma}(l)\right\rangle$ of spin-$\sigma$ in the $l$-th lattice site.
We prepare the initial state as the equal superposition of spin-up and spin-down state in the $0$-th site, which is given as
\begin{equation}
\left|\psi_{I}\right\rangle= \frac{1}{\sqrt{2}} \left( {\begin{matrix}
0 \\
0
\end{matrix}} \right)_{-L} \oplus \left( {\begin{matrix}
1 \\
1
\end{matrix}} \right)_{0} \oplus \left( {\begin{matrix}
0 \\
0
\end{matrix}} \right)_{+L}.
\end{equation}
The input wavepacket is coherently split into two parts at $(-L)$-th and $(+L)$-th lattice sites through the first beam splitter (BS1), which contains a spin-dependent shift and a subsequent $\frac{\pi}{2}$ pulse.
The first shifting operation transfers the initial state $\left|\psi_{I}\right\rangle$ to
\begin{equation}
\left|\psi_{S}\right\rangle= \frac{1}{\sqrt{2}} \left( {\begin{matrix}
e^{i\phi_{1,\uparrow}} \\
0
\end{matrix}} \right)_{-L} \oplus \left( {\begin{matrix}
0 \\
0
\end{matrix}} \right)_{0} \oplus \left( {\begin{matrix}
0 \\
e^{i\phi_{1,\downarrow}}
\end{matrix}} \right)_{+L},
\end{equation}
where $\phi_{1,\uparrow}$ and $\phi_{1,\downarrow}$ are the phases accumulated in the first shifting operation.
The shifting is slow enough that the atom follows the instantaneous Wannier-Stark energy $E_{\uparrow}=\epsilon_{\uparrow} + Fd \nu t/\pi + \hbar \omega_0/2$ in $\left|\uparrow\right\rangle$ and $E_{\downarrow}=\epsilon_{\downarrow} - Fd \nu t/\pi- \hbar \omega_0/2$ in $\left|\downarrow\right\rangle$.
The phases $\phi_{1,\uparrow}$ and $\phi_{1,\downarrow}$ are given as
\begin{eqnarray}
  \phi_{1,\uparrow}= - \frac{1}{\hbar }\left( {\epsilon {_{\uparrow }} + \frac{{\hbar {\omega _0}}}{2} + \frac{{FLd}}{2}} \right){T_s}, \nonumber \\
  \phi_{1,\downarrow}=  - \frac{1}{\hbar }\left( {\epsilon {_{\downarrow }} - \frac{{\hbar {\omega _0}}}{2} - \frac{{FLd}}{2}} \right){T_s},
\end{eqnarray}
where $T_s=l\pi/\nu$ is the shifting time.
Applying a $\frac{\pi}{2}$ pulse operation $e^{-i\hat J_{y}\frac{\pi}{2}}$, the state $\left|\psi\right\rangle_{S}$ is transformed to
\begin{equation}
\left|\psi_{BS1}\right\rangle = \frac{1}{2} \left( {\begin{matrix}
e^{i\theta_{1,\rightarrow}} \\
-e^{i\theta_{1,\rightarrow}}
\end{matrix}} \right)_{-L} \oplus \left( {\begin{matrix}
0 \\
0
\end{matrix}} \right)_{0} \oplus \left( {\begin{matrix}
e^{i\theta_{1,\leftarrow}} \\
e^{i\theta_{1,\leftarrow}}
\end{matrix}} \right)_{+L}. \\
\end{equation}
Here, $\theta_{1,\rightarrow}=\phi_{1,\uparrow}+\phi_{2,\rightarrow}$, $\theta_{1,\leftarrow}=\phi_{1,\downarrow}+\phi_{2,\leftarrow}$, $\phi_{2,\leftarrow}=FLdT_{\pi/2}/{\hbar}$ and $\phi_{2,\rightarrow}=-\phi_{2,\leftarrow}$, where $T_{\pi/2}$ is the lasting time of the $\frac{\pi}{2}$ pulse operation.
After holding the optical lattices still for a time duration $T_h$, the state $\left|\psi_{BS1}\right\rangle$ is transferred to
\begin{equation}
  \left|\psi_{H}\right\rangle=\frac{1}{2} \left( {\begin{matrix}
e^{i\theta_{2,\rightarrow}}e^{-i\xi} \\
-e^{i\theta_{2,\rightarrow}}e^{i\xi}
\end{matrix}} \right)_{-L} \oplus \left( {\begin{matrix}
0 \\
0
\end{matrix}} \right)_{0} \oplus \left( {\begin{matrix}
e^{i\theta_{2,\leftarrow}}e^{-i\xi} \\
e^{i\theta_{2,\leftarrow}}e^{i\xi}
\end{matrix}} \right)_{+L},
\end{equation}
where $\xi={( {{\epsilon {_{\uparrow }} - \epsilon {_{\downarrow }}} + {\hbar {\omega _0}}} ){T_h}}/(2\hbar)$, $\theta_{2,\rightarrow}=\theta_{1,\rightarrow}+\phi_{3,\rightarrow}$, $\theta_{2,\leftarrow}=\theta_{1,\leftarrow}+\phi_{3,\leftarrow}$, $\phi_{3,\leftarrow}=FLdT_{h}/{\hbar}$ and $\phi_{3,\rightarrow}=-\phi_{3,\leftarrow}$.
To recombine the wavepackets in the $(-L)$-th and $(+L)$-th lattice sites into the $0$-th lattice site, we apply the second beam splitter (BS2 - the inverse process of BS1) which contains a $\frac{\pi}{2}$ pulse and a subsequent spin-dependent shift.
Applying a $\frac{\pi}{2}$ pulse, the state $\left|\psi_{H}\right\rangle$ is changed to
\begin{equation}
\left|\psi_{\frac{\pi}{2}}\right\rangle=\frac{1}{\sqrt{2}} \left( {\begin{matrix}
ie^{i\theta_{3,\downarrow}}\sin(\xi) \\
e^{i\theta_{3,\downarrow}}\cos(\xi)
\end{matrix}} \right)_{-L} \oplus \left( {\begin{matrix}
0 \\
0
\end{matrix}} \right)_{0} \oplus \left( {\begin{matrix}
e^{i\theta_{3,\uparrow}}\cos(\xi) \\
ie^{i\theta_{3,\uparrow}}\sin(\xi)
\end{matrix}} \right)_{+L}, \nonumber
\end{equation}
where $\theta_{3,\downarrow}=\theta_{2,\rightarrow}+\phi_{4,\downarrow}$, $\theta_{3,\uparrow}=\theta_{2,\leftarrow}+\phi_{4,\uparrow}$,
$\phi_{4,\uparrow}=FLdT_{\pi/2}/\hbar$, and $\phi_{4,\downarrow}=-\phi_{4,\uparrow}-\pi$.
The second shifting operation transfers the state $\left|\psi_{\pi/2}\right\rangle$ to
\begin{eqnarray}
\left|\psi_{BS2}\right\rangle &=& \frac{1}{\sqrt{2}} \left( {\begin{matrix}
ie^{i\eta}e^{i\theta_{4,\downarrow}}\sin(\xi) \\
0
\end{matrix}} \right)_{-2L} \nonumber \\
&\oplus& \left( {\begin{matrix}
e^{i\theta_{4,\uparrow}}\cos(\xi) \\
e^{i\theta_{4,\downarrow}}\cos(\xi)
\end{matrix}} \right)_{0}  \oplus \left( {\begin{matrix}
0 \\
ie^{-i\eta}e^{i\theta_{4,\uparrow}}\sin(\xi)
\end{matrix}} \right)_{+2L}, \nonumber \\
\end{eqnarray}

where $\eta=-(\epsilon_\uparrow-\epsilon_\downarrow+\hbar\omega_0+FLd)T_s/\hbar$, $\theta_{4,\downarrow}=\theta_{3,\downarrow}+\phi_{5,\downarrow}$, $\theta_{4,\uparrow}=\theta_{3,\uparrow}+\phi_{5,\uparrow}$, $\phi_{5,\uparrow}= - ( {\epsilon {_{\uparrow }} + {\hbar {\omega _0}}/2 - {FLd}/2} ){T_s}/\hbar$ and $\phi_{5,\downarrow}= -( {\epsilon {_{\downarrow }} - {\hbar {\omega _0}}/2 + {FLd}/2} ){T_s}/\hbar$.
Applying the last $\frac{\pi}{2}$ pulse, the final state  $\left|\psi_{F}\right\rangle$ is given as
\begin{eqnarray}
\left|\psi_{F}\right\rangle&=& \frac{1}{2}\left( {\begin{matrix}
ie^{i(\eta-2FLdT_{\pi/2}/\hbar)}e^{i\theta_{4,\downarrow}}\sin(\xi) \\
-ie^{i(\eta-2FLdT_{\pi/2}/\hbar)}e^{i\theta_{4,\downarrow}}\sin(\xi)
\end{matrix}} \right)_{-2L} \nonumber \\
&\oplus& \left( {\begin{matrix}
(e^{i\theta_{4,\uparrow}}+e^{i\theta_{4,\downarrow}})\cos(\xi) \\
(-e^{i\theta_{4,\uparrow}}+e^{i\theta_{4,\downarrow}})\cos(\xi)
\end{matrix}} \right)_{0} \nonumber \\
&\oplus& \left( {\begin{matrix}
ie^{-i(\eta-2FLdT_{\pi/2}/\hbar)}e^{i\theta_{4,\uparrow}}\sin(\xi) \\
ie^{-i(\eta-2FLdT_{\pi/2}/\hbar)}e^{i\theta_{4,\uparrow}}\sin(\xi)
\end{matrix}} \right)_{+2L},
\end{eqnarray}

There are two kinds of spin-population measurements: the local measurement $\langle \hat J_z^{(0)}\rangle$ for the $0$-th lattice site,
and the global measurement $\langle \hat J_z\rangle=\langle \hat J_z^{(0)}\rangle+\langle \hat J_z^{(+2L)}\rangle+\langle \hat J_z^{(-2L)}\rangle$ for all occupied lattice sites. Here, the superscript on the operator denotes the lattice index.
The global measurement of spin population difference is given as,
\begin{eqnarray}
  \langle \hat J_{z}\rangle &=& \langle \hat J_{z}^{(0)}\rangle + \langle \hat J_{z}^{(-2L)}\rangle+ \langle \hat J_{z}^{(+2L)}\rangle \nonumber \\
  &=&\langle \hat J_{z}^{(0)}\rangle = \frac{1}{2}\cos^2(\xi)\cos(\phi),
\end{eqnarray}
which is the same as the local measurement of spin population difference at the $0$-th site.
The total phase is $\phi=\theta_{4,\uparrow}-\theta_{4,\downarrow} =2MgLd(T_s+T_h+2T_{\pi/2})/{\hbar}+\pi$.
The variance in the global measurement is given as
\begin{eqnarray}
  \langle \hat J_{z}\rangle&=&\langle \hat J_{z}^2\rangle-\langle \hat J_{z}\rangle^2 \nonumber \\
  &=&\frac{1}{4}\left[1-\cos^4(\xi)\cos^2(\phi)\right],
  \label{SingleGVariance}
\end{eqnarray}
while the variance in the local measurement is given as
\begin{eqnarray}
\langle \hat J_{z}^{(0)}\rangle&=&\langle (\hat J_{z}^{(0)})^2\rangle-\langle \hat J_{z}^{(0)}\rangle^2 \nonumber \\
  &=&\frac{\cos^2(\xi)}{4}\left[1-\cos^2(\xi)\cos^2(\phi)\right].
  \label{SingleLVariance}
\end{eqnarray}

\section{Multi-particle gravimetry scheme}~\label{Sec4}

In this section, we will generalize the atomic gravimetry interferometry from single particle scheme to multi-particle scheme.
we will show how our scheme works with multi-particle states, such as coherent spin states (CSS's) and spin-squeezed states (SSS's).
Below, we will analytically derive the accumulated phase, the spin population difference and the uncertainty of gravitational acceleration.

\subsection{Phase accumulation}
We assume that $N$ atoms are loaded into the $0$-th lattice site of the first band (the band index $\alpha$ is removed here and after).
The initial state is in superpositions of all possible $n$ particles in $\left|\uparrow\right\rangle_0$ and $(N-n)$ particles in $\left|\downarrow\right\rangle_0$,
\begin{equation}
  \left| \psi_I \right\rangle  = \sum\limits_{n = 0}^N {{c_n}} \left|  \uparrow  \right\rangle _0^{  n}\left|  \downarrow  \right\rangle _0^{  (N - n)},
  \label{psi0}
\end{equation}
where $\left|\sigma\right\rangle_l^n$ denotes $n$ spin-$\sigma$ atoms in the $l$-th lattice site.
Given $c_{N}^n=2^{-N/2} \sqrt{N!/(n!(N-n)!)}$, the initial state is a CSS $|\psi\rangle = 2^{-N/2}(\left|\uparrow\right\rangle_0 +\left|\downarrow\right\rangle_0)^{ N}$.
To beat the SQL, one may squeeze the initial CSS into a phase-sensitive SSS, that is, the noise reduces in the $J_y$-direction and increases in the $J_z$-direction~\cite{Kitagawa1993,Gross2012a,Hosten2016}.

In the first beam splitter (BS1), the spin-up and spin-down components are respectively transported to the $(-L)$-th and $(+L)$-th lattice sites by a coherent spin-dependent shift and then a $\frac{\pi}{2}$ pulse is applied.
Assuming the shift process is adiabatic, the input state evolves to
\begin{equation}
  \left| \psi_{S1}  \right\rangle  = \sum\limits_{n = 0}^N {{c_{N}^n}} e^{i[n\phi_{1,\uparrow}+(N-n)\phi_{1,\downarrow}]} \left|  \uparrow  \right\rangle _{-L}^{ n}\left|  \downarrow  \right\rangle _{+L}^{ (N - n)},
   \label{psi1}
\end{equation}
after the shift.
Here, $T_s = L \pi/\nu$ is the time duration of the shift, and the accumulated phases for the spin-up and spin-down components are respectively given as $\phi_{1,\uparrow}=-  ( {\epsilon {_{\uparrow }} + {\hbar {\omega _0}}/2 + {FLd}/2} ){T_s}/\hbar$ and $\phi_{1,\downarrow}=  - ( {\epsilon {_{\downarrow }} - {\hbar {\omega _0}}/2 - {FLd}/2} ){T_s}/\hbar$.
Then the $\frac{\pi}{2}$ pulse transfers the state $\left| \psi_{I}  \right\rangle$ into
\begin{eqnarray}
 \left|\psi_{BS1}\right\rangle &=&\exp(-i\hat J_y \pi/2) \left|\psi_1\right\rangle\\
 &=&\sum\limits_{n=0}^{N} {{c_{N}^n}e^{i[n\theta_{1,\rightarrow} +(N-n)\theta_{1,\leftarrow}]} { {\left| \rightarrow \right\rangle }_{-L}^{ n}}{ {\left| \leftarrow \right\rangle }_{+L}^{ (N-n)} }}.\nonumber
 \label{psiBS1}
\end{eqnarray}
Here, we denote $\left|\leftarrow \right\rangle_l=(\left|\uparrow\right\rangle_{l}+\left|\downarrow\right\rangle_{l}) /\sqrt{2}$ and $\left|\rightarrow \right\rangle_l=(\left|\uparrow\right\rangle_{l}-\left|\downarrow\right\rangle_{l}) /\sqrt{2}$ at the $l$-th site.
$\theta_{1,\rightarrow}=\phi_{1,\uparrow}+\phi_{2,\rightarrow}$, $\theta_{1,\leftarrow}=\phi_{1,\downarrow}+\phi_{2,\leftarrow}$,  $\phi_{2,\leftarrow}=FLdT_{\pi/2}/\hbar$,  $\phi_{2,\rightarrow}=-\phi_{2,\leftarrow}$, and $T_{\pi/2}$ being the time duration of the $\frac{\pi}{2}$ pulse.

In the holding process, the two parts at different heights will accumulate a relative phase due to the gravity, while the internal spin states will rotate around the $\hat J_z$ axis.
After holding the system still for a time duration $T_h$, the state $\psi_{BS1}$ evolves to
\begin{eqnarray}
 \left|\psi_{H}\right\rangle&=& \sum\limits_{n=0}^{N}\frac{f_n}{2^{N/2}} \left({e^{-i\xi}\left| \uparrow \right\rangle_{-L}-e^{i\xi}\left| \downarrow \right\rangle_{-L}  }\right)^{ n} \nonumber \\
 &\otimes& \left( {e^{-i\xi}{\left| \uparrow \right\rangle_{+L}+e^{i\xi}\left| \downarrow \right\rangle_{+L} }}\right)^{ (N-n)}.
 \label{psiH}
\end{eqnarray}
where $f_n={c_{N}^n}e^{i[n\theta_{2,\rightarrow} +(N-n)\theta_{2,\leftarrow}]}$ with $\theta_{2,\rightarrow}=\theta_{1,\rightarrow}+\phi_{3,\rightarrow}$, $\theta_{2,\leftarrow}=\theta_{1,\leftarrow}+\phi_{3,\leftarrow}$, $\phi_{3,\leftarrow}=FLdT_h/\hbar$, and $\phi_{3,\rightarrow}=-\phi_{3,\leftarrow}$.
In the holding process, the relative phase between $\left|\uparrow\right\rangle$ and $\left|\downarrow\right\rangle$ is given as $2\xi=(\epsilon_{\uparrow}-\epsilon_{\downarrow}+\hbar\omega_0)T_f/\hbar$.
We then apply the second beam splitter (BS2) which consists of a $\frac{\pi}{2}$ pulse and a subsequent shifting process.
After applying the $\frac{\pi}{2}$ pulse in BS2, the state reads as
\begin{eqnarray}
 \left|\psi_{\frac{\pi}{2}}\right\rangle = \sum\limits_{n=0}^{N} && f_n' \left( i\sin(\xi){\left| \uparrow \right\rangle_{-L}+\cos(\xi)\left| \downarrow \right\rangle_{-L}  }\right)^{ n} \nonumber \\
 &&\otimes \left( \cos(\xi){\left| \uparrow \right\rangle_{+L}+i\sin(\xi)\left| \downarrow \right\rangle_{+L}  }\right)^{ (N-n)},
 \label{psipi2}
\end{eqnarray}
where $f_n'=c_N^n e^{i[n\theta_{3,\downarrow}+(N-n)\theta_{3,\uparrow})]}$, $\theta_{3,\downarrow}=\theta_{2,\rightarrow}+\phi_{4,\downarrow}$, $\theta_{3,\uparrow}=\theta_{2,\leftarrow}+\phi_{4,\uparrow}$, $\phi_{4,\uparrow}=FLdT_{\pi/2}/\hbar$, and $\phi_{4,\downarrow}=-\phi_{4,\uparrow}-\pi$.
The second shifting operation is as same as the first one, which transfers the state $|\psi_{\pi/2}\rangle$ into
\begin{eqnarray}
 \left|\psi_{BS2}\right\rangle = && \sum\limits_{n=0}^{N} f_n'' \left( i\sin(\xi)e^{i\eta}{\left| \uparrow \right\rangle_{-2L}+\cos(\xi)\left| \downarrow \right\rangle_{0}  }\right)^{ n} \nonumber \\
 &&\otimes \left( \cos(\xi){\left| \uparrow \right\rangle_{0}+i\sin(\xi)e^{-i\eta}\left| \downarrow \right\rangle_{+2L}  }\right)^{ (N-n)},
 \label{psipi2}
\end{eqnarray}
where $\eta=-(\epsilon_\uparrow-\epsilon_\downarrow +\hbar\omega_0+FLd)T_s/\hbar$, $f_n''=c_{N}^n e^{i[n\theta_{4,\downarrow}+(N-n)\theta_{4,\uparrow})]}$, $\theta_{4,\downarrow}=\theta_{3,\downarrow}+\phi_{5,\downarrow}$ and $\theta_{4,\uparrow}=\theta_{3,\uparrow}+\phi_{5,\uparrow}$, $\phi_{5,\uparrow}= - ( {\epsilon {_{\uparrow }} + {\hbar {\omega _0}}/2 - {FLd}/2} ){T_s}/\hbar$ and $\phi_{5,\downarrow}= -( {\epsilon {_{\downarrow }} - {\hbar {\omega _0}}/2 + {FLd}/2} ){T_s}/\hbar$.
Since the states at different sites will not interfere, when the atoms are recombined into the $0$-th lattice site, applying the last $\frac{\pi}{2}$ pulse, the final state reads as
\begin{eqnarray}
  |\psi_{F}\rangle = \sum\limits_{n,j,k}^{N} && f_n'' d_{n}^j d_{N-n}^k \sin(\xi)^{j+N-n-k}\cos(\xi)^{n-j+k} \nonumber \\
  && \times \left|A_0^j\right\rangle_{-2L} \left|A_{n-j}^k\right\rangle_{0} \left|A_{N-n-k}^0\right\rangle_{+2L}.
  \label{psi5}
\end{eqnarray}
Here, $d_n^j=\sqrt{n!/(j!(n-j)!)}$ and $\left|A_p^q\right\rangle_{l}= (\left|\leftarrow\right\rangle^{ p}  \left| \rightarrow \right\rangle^{q})_l$ denotes $p$ particles in  $\left|\leftarrow\right\rangle$ and $q$ particles in $\left|\rightarrow\right\rangle$ at the $l$-th lattice site.
Obviously, the atoms may occupy four possible states $\{\left|\rightarrow\right\rangle_{-2L}, \left|\leftarrow\right\rangle_{0},  \left|\rightarrow\right\rangle_{0},  \left|\leftarrow\right\rangle_{+2L}\}$ and so that our scheme can be regarded as a four-path interferometry.

\subsection{Spin-population difference}

To derive the expectation values of spin-population difference $\langle \hat J_z\rangle$, we first expand the collective spin operator $\hat J_z$ based on the new single-particle basis $\{ |\rightarrow\rangle,|\leftarrow\rangle\}$, which obey the orthogonal relation, $\langle\rightarrow|\rightarrow\rangle=\langle\leftarrow|\leftarrow\rangle=1$ and $\langle\rightarrow|\leftarrow\rangle=0$.
Considering there are many particles occupied in these two modes, we can expand local state in the fock basis $\{\left|A_p^q\right\rangle_{l} = \left|\leftarrow\right\rangle_{l}^p \left|\rightarrow\right\rangle_{l}^q \}$.
The collective spin operator $\hat J_z$ is given as
\begin{equation}
  \hat J_z=\sum\limits_l \hat J_z^{(l)}=\frac{1}{2}\sum\limits_l\left(\hat a_{\leftarrow,l}^{\dag}\hat a_{\rightarrow,l}+\hat a_{\rightarrow,l}^{\dag}\hat a_{\leftarrow,l}\right).
\end{equation}
Here, The superscripts $l$ on the operator mean that the operator acts on the state at the $l$-th site. $\hat a_{\sigma,l}^{\dag}(\hat a_{\sigma,l})$ annihilates (creates) a boson in the mode $\left|\sigma\right\rangle_l$ ($\sigma=\rightarrow,\leftarrow$) at the $l$-th site.
Thus one can obtain the matrix elements
\begin{equation}
_l \langle A_{p'}^{q'}| \hat J_z^{(l)}|A_p^q\rangle_l = \frac{1}{2}\left(\alpha_p^q\delta_{p',p-1}\delta_{q',q+1}+\alpha_q^p\delta_{p',p+1}\delta_{q',q-1}\right), \nonumber
\end{equation}
and
\begin{eqnarray}
&_l \langle A_{p'}^{q'}|\hat (J_z^{(l)})^2|A_p^q\rangle_l =\frac{1}{4}\delta_{p',p}\delta_{q',q}(\alpha_p^q\alpha_{q+1}^{p-1}+\alpha_q^p\alpha_{p+1}^{q-1}) \nonumber \\ &+\frac{1}{4}\alpha_{p-1}^{q+1}\alpha_p^q\delta_{p',p-2}\delta_{q',q+2}+\frac{1}{4}\alpha_q^p\alpha_{q-1}^{p+1}\delta_{p',p+2}\delta_{q',q-2}, \nonumber
\end{eqnarray}
where $\alpha_p^q=\sqrt{p(q+1)}$. We measure the global mean expectation values of $\hat J_z$, which is given as
\begin{eqnarray}
 \langle\hat J_z\rangle&=&\langle\psi_F|(\hat J_z^{(-2L)}+\hat J_z^{(0)}+\hat J_z^{(+2L)})|\psi_F\rangle \nonumber \\
 &=&\langle\hat J_z^{(-2L)}\rangle+\langle\hat J_z^{(0)}\rangle+\langle\hat J_z^{(+2L)}\rangle.
\end{eqnarray}

We first perform the local operator $\hat J_z^{(-2L)}$ on the final state. $\hat J_z^{(-2L)}$ only acts on $|A_0^j\rangle_{-2L}$ and keeps $|A_{n-j}^k\rangle_{0}$ and $|A_{N-n-k}^0\rangle_{+2L}$ unchanged.
Because $_{-2L}\langle A_0^{j'}|\hat J_z^{-2L}|A_0^{j}\rangle_{-2L}={_{-2L}\langle A_0^{j'}}|A_1^{j-1}\rangle_{-2L}\alpha _{0}^{j}=0$, the mean spin population at the $(-2L)$-th site is zero, $\langle\hat J_z^{(-2L)}\rangle=0$.
Because the states in the $(-2L)$-th and $(+2L)$-th sites have the same status,
the mean spin population at $(+2L)$-th site is also zero, $\langle\hat J_z^{(+2L)}\rangle=0$.

We perform the local operator $J_z^{(0)}$ on the final state, then the mean spin population difference is given as
\begin{widetext}
%
\begin{eqnarray}
  \langle \hat J_z^{(0)}\rangle &=&\frac{1}{2}\sum\limits_{n,j,k} {c_N^n} (c_N^{n - 1})^*{e^{i({\theta _{4, \downarrow }} - {\theta _{4, \uparrow }})}}\alpha _{n - j}^kd_n^jd_{n - 1}^jd_{N - n}^kd_{N - n + 1}^{k + 1}\sin {(\xi )^{2(j + N - n - k)}}\cos {(\xi )^{2(n - j + k)}} \nonumber \\
  &+& \frac{1}{2}\sum\limits_{n,j,k} {c_N^n} (c_N^{n + 1})^*{e^{ - i({\theta _{4, \downarrow }} - {\theta _{4, \uparrow }})}}\alpha _k^{n - j}d_n^jd_{n + 1}^jd_{N - n}^kd_{N - n - 1}^{k - 1}\sin {(\xi )^{2(j + N - n - k)}}\cos {(\xi )^{2(n - j + k)}} \nonumber \\
  &=&\frac{1}{2} \sum\limits_{n,k} {c_N^n} (c_N^{n - 1})^*{e^{i({\theta _{4, \downarrow }} - {\theta _{4, \uparrow }})}}\sqrt {k + 1} \sqrt n d_{N - n}^kd_{N - n + 1}^{k + 1}\sin {(\xi )^{2(N - n - k)}}\cos {(\xi )^{2 + 2k}} \nonumber \\
  &+&\frac{1}{2}\sum\limits_{n,k} {c_N^n} (c_N^{n + 1})^*{e^{ - i({\theta _{4, \downarrow }} - {\theta _{4, \uparrow }})}}\sqrt k \sqrt {n + 1} d_{N - n}^kd_{N - n - 1}^{k - 1}\sin {(\xi )^{2(N - n - k)}}\cos {(\xi )^{2k}} \nonumber \\
  &=&\frac{1}{2}\sum\limits_n {\cos {{(\xi )}^2}\left[\alpha_n^{N-n}c_N^n (c_N^{n - 1})^*{e^{i({\theta _{4, \downarrow }} - {\theta _{4, \uparrow }})}}+\alpha_n^{N-n}(c_N^n)^* c_N^{n - 1}{e^{-i({\theta _{4, \downarrow }} - {\theta _{4, \uparrow }})}}\right] }.
  \end{eqnarray}
\end{widetext}
In the second and third equal signs, we have used the fact that $\sum_{j = 0}^n {\frac{{n!}}{{j!(n - j)!}}\sin {{(\xi )}^{2j}}\cos {{(\xi )}^{2(n - j)}}}  = {( {{{\sin }^2}(\xi ) + {{\cos }^2}(\xi )} )^n} = 1$ for several times, which is also important in the following derivations.
If the initial state satisfies $c_N^{n}=c_N^{N-n}$, then $\sum_{n}\alpha_n^{N-n}c_N^n (c_N^{n - 1})^*$  is a real number.
Our initial CSS and the one-axis twisting SSS satisfy $c_N^{n}=c_N^{N-n}$,
so the above equations can be further simplified as
\begin{equation}
    \langle\hat J_z\rangle= \langle\hat J_z^{(0)}\rangle=\frac{N\mathcal{V}}{2}\cos(\phi).
\end{equation}
It means that both the local and global measurements give the same spin-population difference. Here, $\phi={\theta _{4, \uparrow }} - {\theta _{4, \downarrow }}=2MgLd(T_h +T_s+2T_{\pi/2})/{\hbar}+\pi$, and we define $\mathcal{V}=2\cos^2(\xi)\sum_{n}\alpha_n^{N-n}c_N^n (c_N^{n - 1})^*/N$ as the visibility which reduces to  $\cos^2(\xi)$ for CSS.

\subsection{Uncertainty of gravitational acceleration}

The relative uncertainty of gravitational acceleration is given as $\Delta g/g=\Delta J_z /|\partial \langle J_z\rangle /\partial g|$. To obtain the uncertainty, we need to evaluate $\Delta {J_z}^2 = \langle {{\hat J}_z}^2\rangle-\langle {{\hat J}_z}\rangle^2$. Thus, we will first derive the mean expectation value of $\hat J_z^2$, which is given as
\begin{eqnarray}
\langle {{\hat J}_z}^2\rangle  &=& \langle {\psi _F}|{(\hat J_z^{( - 2L)} + \hat J_z^{(0)} + \hat J_z^{( + 2L)})^2}|{\psi _F}\rangle \nonumber \\
 &=& \langle {(\hat J_z^{( - 2L)})^2}\rangle  + \langle {(\hat J_z^{(0)})^2}\rangle  + \langle {(\hat J_z^{( + 2L)})^2}\rangle ,
\end{eqnarray}
where the crossing terms are all zero. To derive $\langle {{\hat J}_z}^2\rangle$, we separately calculate the mean expectation value of $(\hat J_z^{(\pm 2L)})^2$ and $(\hat J_z^{(0)})^2$ as follows.
\begin{equation}
  \langle(\hat J_z^{(- 2L)})^2\rangle = \frac{{\sin {{(\xi )}^2}}}{4}\sum\limits_n^{} {|c_N^n{|^2}n},
   \label{Jz2lm}
\end{equation}
 which is simplified as $\langle(\hat J_z^{(- 2L)})^2\rangle = \frac{N}{8}{\sin {{(\xi )}^2}}$ for our initial CSS. Similarly,
\begin{equation}
  \langle(\hat J_z^{(+ 2L)})^2\rangle = \frac{{\sin {{(\xi )}^2}}}{4}\sum\limits_n^{} {|c_N^n{|^2}(N - n)},
   \label{Jz2lp}
\end{equation}
which can be also simplified as $\langle(\hat J_z^{(+ 2L)})^2 \rangle= \frac{N}{8}{\sin {{(\xi )}^2}}$ for our initial CSS. At last, we obtain the mean expectation value of $(\hat J_z^{(0)})^2$,
\begin{eqnarray}
  && \langle(\hat J_z^{(0)})^2\rangle        = \frac{{\cos {{(\xi )}^2}N}}{4} + \frac{{\cos {{(\xi )}^4}}}{2}\sum\limits_n^{} {|c_N^n{|^2}n(N - n)} \nonumber \\
  &&+\frac{{\cos {{(\xi )}^4}}}{4}\sum\limits_n  {{e^{i2( {{\theta _{4, \downarrow }} - {\theta _{4, \uparrow }}} )}}c_N^n{{(c_N^{n - 2})}^*}\alpha _{n - 1}^{n - 1}\alpha _{N - n + 1}^{N - n + 1}} \nonumber \\
  &&+\frac{{\cos {{(\xi )}^4}}}{4}\sum\limits_n  {{e^{ - i2( {{\theta _{4, \downarrow }} - {\theta _{4, \uparrow }}} )}}c_N^n{{(c_N^{n + 2})}^*}\alpha _{n + 1}^{n + 1}\alpha _{N - n - 1}^{N - n - 1}}.\nonumber \\
\end{eqnarray}
If the initial state satisfies $c_N^{n}=c_N^{N-n}$, it is easy to prove that $\sum_n  c_N^n{{(c_N^{n - 2})}^*}\alpha _{n - 1}^{n - 1}\alpha _{N - n + 1}^{N - n + 1}$ equals to $\sum_n c_N^n{{(c_N^{n + 2})}^*}\alpha _{n + 1}^{n + 1}\alpha _{N - n - 1}^{N - n - 1}$ $\in  \mathbb{R}$.
Then, the above equation can be simplified as
\begin{eqnarray}
  &\langle(\hat J_z^{(0)})^2\rangle= \frac{{N}}{4}\cos {{(\xi )}^2} + \frac{{\cos {{(\xi )}^4}}}{2}\sum\limits_n^{} {|c_N^n{|^2}n(N - n)} \nonumber \\
&+ \frac{{\cos {{(\xi )}^4}}}{2}\cos(2\phi)\sum\limits_n {c_N^n{{(c_N^{n - 2})}^*}\alpha _{n - 1}^{n - 1}\alpha _{N - n + 1}^{N - n + 1}},
  \label{Jz20}
\end{eqnarray}
which can be further simplified as $\langle(\hat J_z^{(0)})^2\rangle = \frac{N}{4}\cos {(\xi )^2} - \frac{N}{4}\cos {(\xi )^4}{\cos ^2}(\phi ) + \frac{{{N^2}}}{4}\cos {{(\xi )}^4}{\cos ^2}(\phi )$ for our initial CSS.
Combined with Eqs.~\eqref{Jz2lm}, \eqref{Jz2lp} and \eqref{Jz20}, we can obtain
\begin{eqnarray}
  && \langle {{\hat J}_z}^2\rangle = \frac{N}{4}+ \frac{{\cos {{(\xi )}^4}}}{2}\sum\limits_n^{} {|c_N^n{|^2}n(N - n)} \nonumber \\
  && +\frac{{\cos {{(\xi )}^4}}}{2}\cos(2\phi)\sum\limits_n {c_N^n{{(c_N^{n - 2})}^*}\alpha _{n - 1}^{n - 1}\alpha _{N - n + 1}^{N - n + 1}}.
\end{eqnarray}
With the above equation, we only need to know the coefficients of the initial state $C_N^{n}$ to calculate the $\langle {{\hat J}_z}^2\rangle$.

Eventually, the variance of spin population difference for our CSS is given as
\begin{eqnarray}
  \Delta {J_z}^2 = \langle {{\hat J}_z}^2\rangle-\langle {{\hat J}_z}\rangle^2 = \frac{N}{4}\left( {1 - \cos {{(\xi )}^4}{{\cos }^2}(\phi )} \right),
\end{eqnarray}
for the global measurement, and
\begin{eqnarray}
\Delta (J_z^{(0)})^2&=&\langle(\hat J_z^{(0)})^2\rangle- \langle\hat J_z^{(0)}\rangle^2 \nonumber \\
&=& \frac{N}{4}\cos {(\xi )^2} \left[1- \cos {(\xi )^2}{\cos ^2}(\phi )\right],
\end{eqnarray}
for the local measurement.
The above two formula respectively  reduce to Eqs.~\eqref{SingleGVariance} and \eqref{SingleLVariance} when the particle number is $1$.
Obviously, the variance in the local measurement is always not larger than the one in the global measurement.
Actually, the above statement still holds in the case of an input SSS.
The minimal variance is reached at $\xi=(\epsilon_{\uparrow}-\epsilon_{\downarrow}+\hbar\omega_0)T_f /(2\hbar) =n\pi$,
where both local and global measurements give the same variance $\Delta J_z=\Delta J_z^{(0)}$.

To estimate the gravity $g$ with the best precision, besides $\xi=n\pi$, we need to measure the spin-population difference at $\phi=(2n+1)\pi/2$, where both the global and local measurements give the relative uncertainty
\begin{equation}
 \frac{\Delta g}{g}= \frac{\hbar\chi}{2\sqrt{N}MgLd(T_s+T_h+2T_{\pi/2})}.
\end{equation}
Here, $\chi=2\Delta J_z/(\mathcal {V}\sqrt{N})$ is the squeezing parameter, which describes the suppression of phase noises relative to the SQL~\cite{Wineland1992,Gross2012a}.
It has been demonstrated that the squeezing parameter $\chi$ scales with the particle number $N$ as $N^{-1/3}$ for one-axis twisting and $N^{-1/2}$ for two-axis twisting~\cite{Kitagawa1993,Sorensen2001,MA2011}.
To minimize the uncertainty $\Delta g /g$, one may decrease the squeezing parameter, increase the shifting distance or increase the holding time.

\begin{figure}[!htp]
\begin{center}
\includegraphics[width=1\columnwidth]{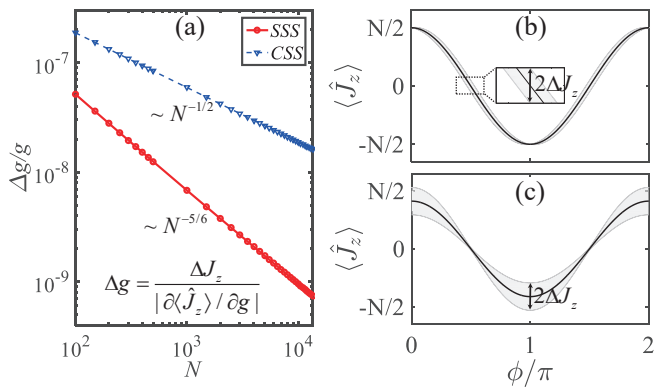}
\end{center}
\caption{(a) The optimal uncertainty $\Delta g/g$ versus the particle number $N$. (b)-(c) The spin-population difference $\langle\hat J_z\rangle$ versus the accumulated phase $\phi$ for for an input CSS and SSS, respectively. The shadow region gives the uncertainty of the spin-population difference. For the input SSS, minimal variance $\Delta J_z$ and maximal slope $|\partial \langle \hat J_z\rangle/\partial \phi|$ are reached at $\phi=(2n+1)\pi/2$.}
\label{phase}
\end{figure}

We estimate the optimal gravity uncertainty for the initial CSS and one-axis-twisted SSS as the particle number changes, see Fig.~\ref{phase}~(a).
The optimal gravity uncertainty scales with the particle number as $N^{-1/2}$ for CSS and $N^{-5/6}$ for the one-axis-twisted SSS~\cite{Kitagawa1993}.
To obtain the optimal gravity uncertainty, one has to minimize the standard variance $\Delta J_z$ and maximize the slope $|\partial \langle \hat J_z\rangle/\partial \phi|$, see Fig.~\ref{phase} (b) for the CSS and (c) for the SSS.
The shadow region ($\langle \hat J_z\rangle\pm\Delta J_z$) denotes the uncertainty of the spin population difference.
The calculations are based on $^{87}R_b$ system with atomic mass $M\approx 1.44 \times {10^{ - 25}}kg$, wavelength $\lambda=7.85\times 10^{-7}m$, recoil energy $E_r=2 \pi^2 \hbar^2 /(M\lambda^2)\approx 2.47 \times {10^{ - 30}}kg \cdot {m^2}/{s^2}$, potential depth $V_{\uparrow}=V_{\downarrow}=100 E_r$, $\hbar \nu=0.5E_r$, $L=50$, $T_{\pi/2}=0.01 ms$, and $T_h\approx 1s$.
Here, the shifting time $T_s=L\pi/\nu\approx 13.4ms$ makes the spin-down and spin-up atoms respectively shift to $\pm L$-th lattices.
Even if the shifting time slightly depart from the ideal value $L\pi/\nu$, our scheme can still work.
The potential depths are large enough to trap more atoms and the driven frequency is small enough to avoid Landau-Zener transition between different Wannier-Stark states.

Now we discuss the importance of the $\frac{\pi}{2}$ pulses in our beam splitters.
Firstly, the $\frac{\pi}{2}$ pulses preclude the background transition energy $\hbar\omega_0$ and the bare on-site energy difference $(\epsilon_\uparrow-\epsilon_{\downarrow})$ in the accumulated phase.
This is because that the switching of spin-up and spin-down automatically cancels out the spin-dependent phases.
Therefore, to extract $g$, we do not need to measure the transition frequency and on-site energy difference at the same time.
Secondly, taking into account the dislocation $\Delta L$ between spin-up and spin-down optical lattices in the holding process, it will directly come into the accumulated phase if the $\frac{\pi}{2}$ pulses are absent, and thus introduce new error sources.
However, by applying the $\frac{\pi}{2}$ pulses, the accumulated phase become relevant to the centers of the spin-up and spin-down mixtures at different heights.
As the dislocation does not change the distance between two centers, it will not contribute to the accumulated phase.
%
%
Similarly, one can use the $\frac{\pi}{2}$ pulses to improve the precision of the digital atom interferometer~\cite{Steffen2012}.

\section{Summary and Discussions}~\label{Sec6}

We present a compact high-precision gravimetry scheme in spin-dependent optical lattices.
Different from the conventional two-path interferometry, our scheme can be treated as a four-path interferometry.
We give an analytical method to analyze the phase accumulation and the gravity measurement variance in our four-path interferometry.
To preclude the spin-dependent energies in the phase accumulation, we introduce an extra $\frac{\pi}{2}$ pulse in each beam splitter.

Our gravimetry scheme can be realized with $^{87}Rb$ atoms in the deep spin-dependent optical lattices aligned along the gravity direction.
The two spin states are chosen as the two hyperfine states of  $^{87}Rb$ atom, $\left|\uparrow\right\rangle\equiv\left|F=2,m_f=-2\right\rangle$ and $\left|\downarrow\right\rangle\equiv\left|F=1,m_f=-1\right\rangle$, which can be resonantly coupled by microwave radiation around $6.8 GHz$~\cite{Mandel2003}.
The wavelength of the optical lattices is tuned to $785$ nm, so that the atoms will feel a spin-dependent optical lattice potential relevant to the light polarization.
With the high-precision and fast optical polarization synthesizer~\cite{Robens2017,Robens2018}, the spin-dependent shifting can be achieved by independently and precisely tuning the spin-up and spin-down optical lattices.
After realizing our scheme with CSS, one may try to utilize SSS to improve the precision of gravity measurement.
To achieve one-axis twisting, one may tune the interaction via Feshabch resonance~\cite{Gross2010} or the spatial overlap between different spin components~\cite{Riedel2010}.

We believe our scheme will open a new era in designing the next-generation high-precision gravimeter.
In contrast to the quantum gravimetry via free fall~\cite{Kasevich1992,Peters1999,Hu2013,Dickerson2013}, our scheme is more compact and portable.
In contrast to the gravimetry via Bloch oscillations in optical lattices~\cite{Ferrari2006}, the spin squeezing may be utilized to beat the standard quantum limit.
In addition to precision gravity measurement, our analytical method for the sensitivity can be widely used in multi-parameter estimation and multi-path interferometry, and the BS's in our scheme can also be applied to the digital atomic interferometry~\cite{Steffen2012} for improving its measurement precision.
Moreover, our study may also advance further experimental studies in gravitational wave detection~\cite{Dimopoulos}, Casimir-Polder force~\cite{Sukenik}, and blackbody radiation induced force~\cite{Haslinger}, etc.

\begin{acknowledgments}
The authors thank Prof. Jun Luo and Prof. Zhongkun Hu for their supports for initiating the research project on precision gravity measurement. This work was supported by the National Natural Science Foundation of China (NNSFC) under Grants No. 11874434, 11574405 and 11704420. Y. Ke was partially supported by International Postdoctoral Exchange Fellowship Program (No. 20180052). J.H. was partially supported by National Postdoctoral Program for Innovative Talents of China (BX201600198).
\end{acknowledgments}

\bibliography{gravimetry}

\begin{thebibliography}{36}%
\makeatletter
\providecommand \@ifxundefined [1]{%
 \@ifx{#1\undefined}
}%
\providecommand \@ifnum [1]{%
 \ifnum #1\expandafter \@firstoftwo
 \else \expandafter \@secondoftwo
 \fi
}%
\providecommand \@ifx [1]{%
 \ifx #1\expandafter \@firstoftwo
 \else \expandafter \@secondoftwo
 \fi
}%
\providecommand \natexlab [1]{#1}%
\providecommand \enquote  [1]{``#1''}%
\providecommand \bibnamefont  [1]{#1}%
\providecommand \bibfnamefont [1]{#1}%
\providecommand \citenamefont [1]{#1}%
\providecommand \href@noop [0]{\@secondoftwo}%
\providecommand \href [0]{\begingroup \@sanitize@url \@href}%
\providecommand \@href[1]{\@@startlink{#1}\@@href}%
\providecommand \@@href[1]{\endgroup#1\@@endlink}%
\providecommand \@sanitize@url [0]{\catcode `\\12\catcode `\$12\catcode
  `\&12\catcode `\#12\catcode `\^12\catcode `\_12\catcode `\%12\relax}%
\providecommand \@@startlink[1]{}%
\providecommand \@@endlink[0]{}%
\providecommand \url  [0]{\begingroup\@sanitize@url \@url }%
\providecommand \@url [1]{\endgroup\@href {#1}{\urlprefix }}%
\providecommand \urlprefix  [0]{URL }%
\providecommand \Eprint [0]{\href }%
\providecommand \doibase [0]{http://dx.doi.org/}%
\providecommand \selectlanguage [0]{\@gobble}%
\providecommand \bibinfo  [0]{\@secondoftwo}%
\providecommand \bibfield  [0]{\@secondoftwo}%
\providecommand \translation [1]{[#1]}%
\providecommand \BibitemOpen [0]{}%
\providecommand \bibitemStop [0]{}%
\providecommand \bibitemNoStop [0]{.\EOS\space}%
\providecommand \EOS [0]{\spacefactor3000\relax}%
\providecommand \BibitemShut  [1]{\csname bibitem#1\endcsname}%
\let\auto@bib@innerbib\@empty
\bibitem [{\citenamefont {de~Angelis}\ \emph {et~al.}(2009)\citenamefont
  {de~Angelis}, \citenamefont {Bertoldi}, \citenamefont {Cacciapuoti},
  \citenamefont {Giorgini}, \citenamefont {Lamporesi}, \citenamefont
  {Prevedelli}, \citenamefont {Saccorotti}, \citenamefont {Sorrentino},\ and\
  \citenamefont {Tino}}]{Angelis2009}%
  \BibitemOpen
  \bibfield  {author} {\bibinfo {author} {\bibfnamefont {M.}~\bibnamefont
  {de~Angelis}}, \bibinfo {author} {\bibfnamefont {A.}~\bibnamefont
  {Bertoldi}}, \bibinfo {author} {\bibfnamefont {L.}~\bibnamefont
  {Cacciapuoti}}, \bibinfo {author} {\bibfnamefont {A.}~\bibnamefont
  {Giorgini}}, \bibinfo {author} {\bibfnamefont {G.}~\bibnamefont {Lamporesi}},
  \bibinfo {author} {\bibfnamefont {M.}~\bibnamefont {Prevedelli}}, \bibinfo
  {author} {\bibfnamefont {G.}~\bibnamefont {Saccorotti}}, \bibinfo {author}
  {\bibfnamefont {F~.}\ \bibnamefont {Sorrentino}}, \ and\ \bibinfo {author}
  {\bibfnamefont {G.~M.}\ \bibnamefont {Tino}},\ }\bibfield  {title} {\enquote
  {\bibinfo {title} {Precision gravimetry with atomic sensors},}\ }\href
  {http://stacks.iop.org/0957-0233/20/i=2/a=022001} {\bibfield  {journal}
  {\bibinfo  {journal} {Measurement Science and Technology}\ }\textbf {\bibinfo
  {volume} {20}},\ \bibinfo {pages} {022001} (\bibinfo {year}
  {2009})}\BibitemShut {NoStop}%
\bibitem [{\citenamefont {Richter}\ \emph {et~al.}(1995)\citenamefont
  {Richter}, \citenamefont {Wilmes},\ and\ \citenamefont
  {Nowak}}]{Richter1995}%
  \BibitemOpen
  \bibfield  {author} {\bibinfo {author} {\bibfnamefont {B.}~\bibnamefont
  {Richter}}, \bibinfo {author} {\bibfnamefont {H.}~\bibnamefont {Wilmes}}, \
  and\ \bibinfo {author} {\bibfnamefont {I.}~\bibnamefont {Nowak}},\ }\bibfield
   {title} {\enquote {\bibinfo {title} {The frankfurt calibration system for
  relative gravimeters},}\ }\href
  {http://stacks.iop.org/0026-1394/32/i=3/a=010} {\bibfield  {journal}
  {\bibinfo  {journal} {Metrologia}\ }\textbf {\bibinfo {volume} {32}},\
  \bibinfo {pages} {217} (\bibinfo {year} {1995})}\BibitemShut {NoStop}%
\bibitem [{\citenamefont {Niebauer}\ \emph {et~al.}(1995)\citenamefont
  {Niebauer}, \citenamefont {Sasagawa}, \citenamefont {Faller}, \citenamefont
  {Hilt},\ and\ \citenamefont {Klopping}}]{Niebauer1995}%
  \BibitemOpen
  \bibfield  {author} {\bibinfo {author} {\bibfnamefont {T.~M.}\ \bibnamefont
  {Niebauer}}, \bibinfo {author} {\bibfnamefont {G.~S.}\ \bibnamefont
  {Sasagawa}}, \bibinfo {author} {\bibfnamefont {J.~E.}\ \bibnamefont
  {Faller}}, \bibinfo {author} {\bibfnamefont {R.}~\bibnamefont {Hilt}}, \ and\
  \bibinfo {author} {\bibfnamefont {F.}~\bibnamefont {Klopping}},\ }\bibfield
  {title} {\enquote {\bibinfo {title} {A new generation of absolute
  gravimeters},}\ }\href {http://stacks.iop.org/0026-1394/32/i=3/a=004}
  {\bibfield  {journal} {\bibinfo  {journal} {Metrologia}\ }\textbf {\bibinfo
  {volume} {32}},\ \bibinfo {pages} {159} (\bibinfo {year} {1995})}\BibitemShut
  {NoStop}%
\bibitem [{\citenamefont {Goodkind}(1999)}]{John1999}%
  \BibitemOpen
  \bibfield  {author} {\bibinfo {author} {\bibfnamefont {John~M.}\ \bibnamefont
  {Goodkind}},\ }\bibfield  {title} {\enquote {\bibinfo {title} {The
  superconducting gravimeter},}\ }\href {\doibase 10.1063/1.1150092} {\bibfield
   {journal} {\bibinfo  {journal} {Review of Scientific Instruments}\ }\textbf
  {\bibinfo {volume} {70}},\ \bibinfo {pages} {4131--4152} (\bibinfo {year}
  {1999})}\BibitemShut {NoStop}%
\bibitem [{\citenamefont {et. al.}(2012)}]{Jiang2012}%
  \BibitemOpen
  \bibfield  {author} {\bibinfo {author} {\bibfnamefont {Z.~Jiang}\
  \bibnamefont {et. al.}},\ }\bibfield  {title} {\enquote {\bibinfo {title}
  {The 8th international comparison of absolute gravimeters 2009: the first key
  comparison (ccm.g-k1) in the field of absolute gravimetry},}\ }\href
  {http://stacks.iop.org/0026-1394/49/i=6/a=666} {\bibfield  {journal}
  {\bibinfo  {journal} {Metrologia}\ }\textbf {\bibinfo {volume} {49}},\
  \bibinfo {pages} {666} (\bibinfo {year} {2012})}\BibitemShut {NoStop}%
\bibitem [{\citenamefont {Kasevich}\ and\ \citenamefont
  {Chu}(1992)}]{Kasevich1992}%
  \BibitemOpen
  \bibfield  {author} {\bibinfo {author} {\bibfnamefont {M.}~\bibnamefont
  {Kasevich}}\ and\ \bibinfo {author} {\bibfnamefont {S.}~\bibnamefont {Chu}},\
  }\bibfield  {title} {\enquote {\bibinfo {title} {Measurement of the
  gravitational acceleration of an atom with a light-pulse atom
  interferometer},}\ }\href {\doibase 10.1007/BF00325375} {\bibfield  {journal}
  {\bibinfo  {journal} {Appl. Phys. B}\ }\textbf {\bibinfo {volume} {54}},\
  \bibinfo {pages} {321--332} (\bibinfo {year} {1992})}\BibitemShut {NoStop}%
\bibitem [{\citenamefont {Peters}\ \emph {et~al.}(1999)\citenamefont {Peters},
  \citenamefont {Chung},\ and\ \citenamefont {Chu}}]{Peters1999}%
  \BibitemOpen
  \bibfield  {author} {\bibinfo {author} {\bibfnamefont {A.}~\bibnamefont
  {Peters}}, \bibinfo {author} {\bibfnamefont {K.~Y.}\ \bibnamefont {Chung}}, \
  and\ \bibinfo {author} {\bibfnamefont {S.}~\bibnamefont {Chu}},\ }\bibfield
  {title} {\enquote {\bibinfo {title} {Measurement of gravitational
  acceleration by dropping atoms},}\ }\href {\doibase 10.1038/23655} {\bibfield
   {journal} {\bibinfo  {journal} {Nature}\ }\textbf {\bibinfo {volume}
  {400}},\ \bibinfo {pages} {849--852} (\bibinfo {year} {1999})}\BibitemShut
  {NoStop}%
\bibitem [{\citenamefont {Hu}\ \emph {et~al.}(2013)\citenamefont {Hu},
  \citenamefont {Sun}, \citenamefont {Duan}, \citenamefont {Zhou},
  \citenamefont {Chen}, \citenamefont {Zhan}, \citenamefont {Zhang},\ and\
  \citenamefont {Luo}}]{Hu2013}%
  \BibitemOpen
  \bibfield  {author} {\bibinfo {author} {\bibfnamefont {Z.-K.}\ \bibnamefont
  {Hu}}, \bibinfo {author} {\bibfnamefont {B.-L.}\ \bibnamefont {Sun}},
  \bibinfo {author} {\bibfnamefont {X.-C.}\ \bibnamefont {Duan}}, \bibinfo
  {author} {\bibfnamefont {M.-K.}\ \bibnamefont {Zhou}}, \bibinfo {author}
  {\bibfnamefont {L.-L.}\ \bibnamefont {Chen}}, \bibinfo {author}
  {\bibfnamefont {S.}~\bibnamefont {Zhan}}, \bibinfo {author} {\bibfnamefont
  {Q.-Z.}\ \bibnamefont {Zhang}}, \ and\ \bibinfo {author} {\bibfnamefont
  {J.}~\bibnamefont {Luo}},\ }\bibfield  {title} {\enquote {\bibinfo {title}
  {Demonstration of an ultrahigh-sensitivity atom-interferometry absolute
  gravimeter},}\ }\href {\doibase 10.1103/PhysRevA.88.043610} {\bibfield
  {journal} {\bibinfo  {journal} {Phys. Rev. A}\ }\textbf {\bibinfo {volume}
  {88}},\ \bibinfo {pages} {043610} (\bibinfo {year} {2013})}\BibitemShut
  {NoStop}%
\bibitem [{\citenamefont {Dickerson}\ \emph {et~al.}(2013)\citenamefont
  {Dickerson}, \citenamefont {Hogan}, \citenamefont {Sugarbaker}, \citenamefont
  {Johnson},\ and\ \citenamefont {Kasevich}}]{Dickerson2013}%
  \BibitemOpen
  \bibfield  {author} {\bibinfo {author} {\bibfnamefont {S.~M.}\ \bibnamefont
  {Dickerson}}, \bibinfo {author} {\bibfnamefont {J.~M.}\ \bibnamefont
  {Hogan}}, \bibinfo {author} {\bibfnamefont {A.}~\bibnamefont {Sugarbaker}},
  \bibinfo {author} {\bibfnamefont {D.~M.~S.}\ \bibnamefont {Johnson}}, \ and\
  \bibinfo {author} {\bibfnamefont {M.~A.}\ \bibnamefont {Kasevich}},\
  }\bibfield  {title} {\enquote {\bibinfo {title} {Multiaxis inertial sensing
  with long-time point source atom interferometry},}\ }\href {\doibase
  10.1103/PhysRevLett.111.083001} {\bibfield  {journal} {\bibinfo  {journal}
  {Phys. Rev. Lett.}\ }\textbf {\bibinfo {volume} {111}},\ \bibinfo {pages}
  {083001} (\bibinfo {year} {2013})}\BibitemShut {NoStop}%
\bibitem [{\citenamefont {Ferrari}\ \emph {et~al.}(2006)\citenamefont
  {Ferrari}, \citenamefont {Poli}, \citenamefont {Sorrentino},\ and\
  \citenamefont {Tino}}]{Ferrari2006}%
  \BibitemOpen
  \bibfield  {author} {\bibinfo {author} {\bibfnamefont {G.}~\bibnamefont
  {Ferrari}}, \bibinfo {author} {\bibfnamefont {N.}~\bibnamefont {Poli}},
  \bibinfo {author} {\bibfnamefont {F.}~\bibnamefont {Sorrentino}}, \ and\
  \bibinfo {author} {\bibfnamefont {G.~M.}\ \bibnamefont {Tino}},\ }\bibfield
  {title} {\enquote {\bibinfo {title} {Long-lived bloch oscillations with
  bosonic sr atoms and application to gravity measurement at the micrometer
  scale},}\ }\href {\doibase 10.1103/PhysRevLett.97.060402} {\bibfield
  {journal} {\bibinfo  {journal} {Phys. Rev. Lett.}\ }\textbf {\bibinfo
  {volume} {97}},\ \bibinfo {pages} {060402} (\bibinfo {year}
  {2006})}\BibitemShut {NoStop}%
\bibitem [{\citenamefont {Hughes}\ \emph {et~al.}(2009)\citenamefont {Hughes},
  \citenamefont {Burke},\ and\ \citenamefont {Sackett}}]{Hughes2009}%
  \BibitemOpen
  \bibfield  {author} {\bibinfo {author} {\bibfnamefont {K.~J.}\ \bibnamefont
  {Hughes}}, \bibinfo {author} {\bibfnamefont {J.~H.~T.}\ \bibnamefont
  {Burke}}, \ and\ \bibinfo {author} {\bibfnamefont {C.~A.}\ \bibnamefont
  {Sackett}},\ }\bibfield  {title} {\enquote {\bibinfo {title} {Suspension of
  atoms using optical pulses, and application to gravimetry},}\ }\href
  {\doibase 10.1103/PhysRevLett.102.150403} {\bibfield  {journal} {\bibinfo
  {journal} {Phys. Rev. Lett.}\ }\textbf {\bibinfo {volume} {102}},\ \bibinfo
  {pages} {150403} (\bibinfo {year} {2009})}\BibitemShut {NoStop}%
\bibitem [{\citenamefont {Beaufils}\ \emph {et~al.}(2011)\citenamefont
  {Beaufils}, \citenamefont {Tackmann}, \citenamefont {Wang}, \citenamefont
  {Pelle}, \citenamefont {Pelisson}, \citenamefont {Wolf},\ and\ \citenamefont
  {dos Santos}}]{Beaufils2011}%
  \BibitemOpen
  \bibfield  {author} {\bibinfo {author} {\bibfnamefont {Q.}~\bibnamefont
  {Beaufils}}, \bibinfo {author} {\bibfnamefont {G.}~\bibnamefont {Tackmann}},
  \bibinfo {author} {\bibfnamefont {X.}~\bibnamefont {Wang}}, \bibinfo {author}
  {\bibfnamefont {B.}~\bibnamefont {Pelle}}, \bibinfo {author} {\bibfnamefont
  {S.}~\bibnamefont {Pelisson}}, \bibinfo {author} {\bibfnamefont
  {P.}~\bibnamefont {Wolf}}, \ and\ \bibinfo {author} {\bibfnamefont
  {F.~Pereira}\ \bibnamefont {dos Santos}},\ }\bibfield  {title} {\enquote
  {\bibinfo {title} {Laser controlled tunneling in a vertical optical
  lattice},}\ }\href {\doibase 10.1103/PhysRevLett.106.213002} {\bibfield
  {journal} {\bibinfo  {journal} {Phys. Rev. Lett.}\ }\textbf {\bibinfo
  {volume} {106}},\ \bibinfo {pages} {213002} (\bibinfo {year}
  {2011})}\BibitemShut {NoStop}%
\bibitem [{\citenamefont {Charri\`ere}\ \emph {et~al.}(2012)\citenamefont
  {Charri\`ere}, \citenamefont {Cadoret}, \citenamefont {Zahzam}, \citenamefont
  {Bidel},\ and\ \citenamefont {Bresson}}]{Cadoret2012}%
  \BibitemOpen
  \bibfield  {author} {\bibinfo {author} {\bibfnamefont {R.}~\bibnamefont
  {Charri\`ere}}, \bibinfo {author} {\bibfnamefont {M.}~\bibnamefont
  {Cadoret}}, \bibinfo {author} {\bibfnamefont {N.}~\bibnamefont {Zahzam}},
  \bibinfo {author} {\bibfnamefont {Y.}~\bibnamefont {Bidel}}, \ and\ \bibinfo
  {author} {\bibfnamefont {A.}~\bibnamefont {Bresson}},\ }\bibfield  {title}
  {\enquote {\bibinfo {title} {Local gravity measurement with the combination
  of atom interferometry and bloch oscillations},}\ }\href {\doibase
  10.1103/PhysRevA.85.013639} {\bibfield  {journal} {\bibinfo  {journal} {Phys.
  Rev. A}\ }\textbf {\bibinfo {volume} {85}},\ \bibinfo {pages} {013639}
  (\bibinfo {year} {2012})}\BibitemShut {NoStop}%
\bibitem [{\citenamefont {Andia}\ \emph {et~al.}(2013)\citenamefont {Andia},
  \citenamefont {Jannin}, \citenamefont {Nez}, \citenamefont {Biraben},
  \citenamefont {Guellati-Kh\'elifa},\ and\ \citenamefont
  {Clad\'e}}]{Andia2013}%
  \BibitemOpen
  \bibfield  {author} {\bibinfo {author} {\bibfnamefont {M.}~\bibnamefont
  {Andia}}, \bibinfo {author} {\bibfnamefont {R.}~\bibnamefont {Jannin}},
  \bibinfo {author} {\bibfnamefont {F.\ifmmode
  \mbox{\c{c}}\else~\c{c}\fi{}ois}\ \bibnamefont {Nez}}, \bibinfo {author}
  {\bibfnamefont {F.\ifmmode \mbox{\c{c}}\else~\c{c}\fi{}ois}\ \bibnamefont
  {Biraben}}, \bibinfo {author} {\bibfnamefont {S.}~\bibnamefont
  {Guellati-Kh\'elifa}}, \ and\ \bibinfo {author} {\bibfnamefont
  {P.}~\bibnamefont {Clad\'e}},\ }\bibfield  {title} {\enquote {\bibinfo
  {title} {Compact atomic gravimeter based on a pulsed and accelerated optical
  lattice},}\ }\href {\doibase 10.1103/PhysRevA.88.031605} {\bibfield
  {journal} {\bibinfo  {journal} {Phys. Rev. A}\ }\textbf {\bibinfo {volume}
  {88}},\ \bibinfo {pages} {031605} (\bibinfo {year} {2013})}\BibitemShut
  {NoStop}%
\bibitem [{\citenamefont {Hamilton}\ \emph {et~al.}(2015)\citenamefont
  {Hamilton}, \citenamefont {Jaffe}, \citenamefont {Brown}, \citenamefont
  {Maisenbacher}, \citenamefont {Estey},\ and\ \citenamefont
  {M\"uller}}]{Hamilton2015}%
  \BibitemOpen
  \bibfield  {author} {\bibinfo {author} {\bibfnamefont {P.}~\bibnamefont
  {Hamilton}}, \bibinfo {author} {\bibfnamefont {M.}~\bibnamefont {Jaffe}},
  \bibinfo {author} {\bibfnamefont {J.~M.}\ \bibnamefont {Brown}}, \bibinfo
  {author} {\bibfnamefont {L.}~\bibnamefont {Maisenbacher}}, \bibinfo {author}
  {\bibfnamefont {B.}~\bibnamefont {Estey}}, \ and\ \bibinfo {author}
  {\bibfnamefont {H.}~\bibnamefont {M\"uller}},\ }\bibfield  {title} {\enquote
  {\bibinfo {title} {Atom interferometry in an optical cavity},}\ }\href
  {\doibase 10.1103/PhysRevLett.114.100405} {\bibfield  {journal} {\bibinfo
  {journal} {Phys. Rev. Lett.}\ }\textbf {\bibinfo {volume} {114}},\ \bibinfo
  {pages} {100405} (\bibinfo {year} {2015})}\BibitemShut {NoStop}%
\bibitem [{\citenamefont {Abend}\ \emph {et~al.}(2016)\citenamefont {Abend},
  \citenamefont {Gebbe}, \citenamefont {Gersemann}, \citenamefont {Ahlers},
  \citenamefont {M\"untinga}, \citenamefont {Giese}, \citenamefont {Gaaloul},
  \citenamefont {Schubert}, \citenamefont {L\"ammerzahl}, \citenamefont
  {Ertmer}, \citenamefont {Schleich},\ and\ \citenamefont {Rasel}}]{Abend2016}%
  \BibitemOpen
  \bibfield  {author} {\bibinfo {author} {\bibfnamefont {S.}~\bibnamefont
  {Abend}}, \bibinfo {author} {\bibfnamefont {M.}~\bibnamefont {Gebbe}},
  \bibinfo {author} {\bibfnamefont {M.}~\bibnamefont {Gersemann}}, \bibinfo
  {author} {\bibfnamefont {H.}~\bibnamefont {Ahlers}}, \bibinfo {author}
  {\bibfnamefont {H.}~\bibnamefont {M\"untinga}}, \bibinfo {author}
  {\bibfnamefont {E.}~\bibnamefont {Giese}}, \bibinfo {author} {\bibfnamefont
  {N.}~\bibnamefont {Gaaloul}}, \bibinfo {author} {\bibfnamefont
  {C.}~\bibnamefont {Schubert}}, \bibinfo {author} {\bibfnamefont
  {C.}~\bibnamefont {L\"ammerzahl}}, \bibinfo {author} {\bibfnamefont
  {W.}~\bibnamefont {Ertmer}}, \bibinfo {author} {\bibfnamefont {W.~P.}\
  \bibnamefont {Schleich}}, \ and\ \bibinfo {author} {\bibfnamefont {E.~M.}\
  \bibnamefont {Rasel}},\ }\bibfield  {title} {\enquote {\bibinfo {title}
  {Atom-chip fountain gravimeter},}\ }\href {\doibase
  10.1103/PhysRevLett.117.203003} {\bibfield  {journal} {\bibinfo  {journal}
  {Phys. Rev. Lett.}\ }\textbf {\bibinfo {volume} {117}},\ \bibinfo {pages}
  {203003} (\bibinfo {year} {2016})}\BibitemShut {NoStop}%
\bibitem [{\citenamefont {Wannier}(1960)}]{Wannier1960}%
  \BibitemOpen
  \bibfield  {author} {\bibinfo {author} {\bibfnamefont {G.~H.}\ \bibnamefont
  {Wannier}},\ }\bibfield  {title} {\enquote {\bibinfo {title} {Wave functions
  and effective hamiltonian for bloch electrons in an electric field},}\ }\href
  {\doibase 10.1103/PhysRev.117.432} {\bibfield  {journal} {\bibinfo  {journal}
  {Phys. Rev.}\ }\textbf {\bibinfo {volume} {117}},\ \bibinfo {pages}
  {432--439} (\bibinfo {year} {1960})}\BibitemShut {NoStop}%
\bibitem [{\citenamefont {Poli}\ \emph {et~al.}(2011)\citenamefont {Poli},
  \citenamefont {Wang}, \citenamefont {Tarallo}, \citenamefont {Alberti},
  \citenamefont {Prevedelli},\ and\ \citenamefont {Tino}}]{Poli2011}%
  \BibitemOpen
  \bibfield  {author} {\bibinfo {author} {\bibfnamefont {N.}~\bibnamefont
  {Poli}}, \bibinfo {author} {\bibfnamefont {F.-Y.}\ \bibnamefont {Wang}},
  \bibinfo {author} {\bibfnamefont {M.~G.}\ \bibnamefont {Tarallo}}, \bibinfo
  {author} {\bibfnamefont {A.}~\bibnamefont {Alberti}}, \bibinfo {author}
  {\bibfnamefont {M.}~\bibnamefont {Prevedelli}}, \ and\ \bibinfo {author}
  {\bibfnamefont {G.~M.}\ \bibnamefont {Tino}},\ }\bibfield  {title} {\enquote
  {\bibinfo {title} {Precision measurement of gravity with cold atoms in an
  optical lattice and comparison with a classical gravimeter},}\ }\href
  {\doibase 10.1103/PhysRevLett.106.038501} {\bibfield  {journal} {\bibinfo
  {journal} {Phys. Rev. Lett.}\ }\textbf {\bibinfo {volume} {106}},\ \bibinfo
  {pages} {038501} (\bibinfo {year} {2011})}\BibitemShut {NoStop}%
\bibitem [{\citenamefont {Tarallo}\ \emph {et~al.}(2012)\citenamefont
  {Tarallo}, \citenamefont {Alberti}, \citenamefont {Poli}, \citenamefont
  {Chiofalo}, \citenamefont {Wang},\ and\ \citenamefont {Tino}}]{Tarallo2012}%
  \BibitemOpen
  \bibfield  {author} {\bibinfo {author} {\bibfnamefont {M.~G.}\ \bibnamefont
  {Tarallo}}, \bibinfo {author} {\bibfnamefont {A.}~\bibnamefont {Alberti}},
  \bibinfo {author} {\bibfnamefont {N.}~\bibnamefont {Poli}}, \bibinfo {author}
  {\bibfnamefont {M.~L.}\ \bibnamefont {Chiofalo}}, \bibinfo {author}
  {\bibfnamefont {F.-Y.}\ \bibnamefont {Wang}}, \ and\ \bibinfo {author}
  {\bibfnamefont {G.~M.}\ \bibnamefont {Tino}},\ }\bibfield  {title} {\enquote
  {\bibinfo {title} {Delocalization-enhanced bloch oscillations and driven
  resonant tunneling in optical lattices for precision force measurements},}\
  }\href {\doibase 10.1103/PhysRevA.86.033615} {\bibfield  {journal} {\bibinfo
  {journal} {Phys. Rev. A}\ }\textbf {\bibinfo {volume} {86}},\ \bibinfo
  {pages} {033615} (\bibinfo {year} {2012})}\BibitemShut {NoStop}%
\bibitem [{\citenamefont {Pelle}\ \emph {et~al.}(2013)\citenamefont {Pelle},
  \citenamefont {Hilico}, \citenamefont {Tackmann}, \citenamefont {Beaufils},\
  and\ \citenamefont {P.~d. Santos}}]{Pelle2013}%
  \BibitemOpen
  \bibfield  {author} {\bibinfo {author} {\bibfnamefont {B.}~\bibnamefont
  {Pelle}}, \bibinfo {author} {\bibfnamefont {A.}~\bibnamefont {Hilico}},
  \bibinfo {author} {\bibfnamefont {G.}~\bibnamefont {Tackmann}}, \bibinfo
  {author} {\bibfnamefont {Q.}~\bibnamefont {Beaufils}}, \ and\ \bibinfo
  {author} {\bibfnamefont {F.}~\bibnamefont {P.~d. Santos}},\ }\bibfield
  {title} {\enquote {\bibinfo {title} {State-labeling wannier-stark atomic
  interferometers},}\ }\href {\doibase 10.1103/PhysRevA.87.023601} {\bibfield
  {journal} {\bibinfo  {journal} {Phys. Rev. A}\ }\textbf {\bibinfo {volume}
  {87}},\ \bibinfo {pages} {023601} (\bibinfo {year} {2013})}\BibitemShut
  {NoStop}%
\bibitem [{\citenamefont {Mandel}\ \emph {et~al.}(2003)\citenamefont {Mandel},
  \citenamefont {Greiner}, \citenamefont {Widera}, \citenamefont {Rom},
  \citenamefont {Hansch},\ and\ \citenamefont {Bloch}}]{Mandel2003}%
  \BibitemOpen
  \bibfield  {author} {\bibinfo {author} {\bibfnamefont {Olaf}\ \bibnamefont
  {Mandel}}, \bibinfo {author} {\bibfnamefont {Markus}\ \bibnamefont
  {Greiner}}, \bibinfo {author} {\bibfnamefont {Artur}\ \bibnamefont {Widera}},
  \bibinfo {author} {\bibfnamefont {Tim}\ \bibnamefont {Rom}}, \bibinfo
  {author} {\bibfnamefont {Theodor~W.}\ \bibnamefont {Hansch}}, \ and\ \bibinfo
  {author} {\bibfnamefont {Immanuel}\ \bibnamefont {Bloch}},\ }\bibfield
  {title} {\enquote {\bibinfo {title} {Controlled collisions for multi-particle
  entanglement of optically trapped atoms},}\ }\href {\doibase
  10.1038/nature02008} {\bibfield  {journal} {\bibinfo  {journal} {Nature}\
  }\textbf {\bibinfo {volume} {425}},\ \bibinfo {pages} {937--940} (\bibinfo
  {year} {2003})}\BibitemShut {NoStop}%
\bibitem [{\citenamefont {Steffen}\ \emph {et~al.}(2012)\citenamefont
  {Steffen}, \citenamefont {Alberti}, \citenamefont {Alt}, \citenamefont
  {Belmechri}, \citenamefont {Hild}, \citenamefont {Karski}, \citenamefont
  {Widera},\ and\ \citenamefont {Meschede}}]{Steffen2012}%
  \BibitemOpen
  \bibfield  {author} {\bibinfo {author} {\bibfnamefont {A.}~\bibnamefont
  {Steffen}}, \bibinfo {author} {\bibfnamefont {A.}~\bibnamefont {Alberti}},
  \bibinfo {author} {\bibfnamefont {W.}~\bibnamefont {Alt}}, \bibinfo {author}
  {\bibfnamefont {N.}~\bibnamefont {Belmechri}}, \bibinfo {author}
  {\bibfnamefont {S.}~\bibnamefont {Hild}}, \bibinfo {author} {\bibfnamefont
  {M.}~\bibnamefont {Karski}}, \bibinfo {author} {\bibfnamefont {Artur}\
  \bibnamefont {Widera}}, \ and\ \bibinfo {author} {\bibfnamefont
  {D.}~\bibnamefont {Meschede}},\ }\bibfield  {title} {\enquote {\bibinfo
  {title} {Digital atom interferometer with single particle control on a
  discretized space-time geometry},}\ }\href {\doibase 10.1073/pnas.1204285109}
  {\bibfield  {journal} {\bibinfo  {journal} {PNAS}\ }\textbf {\bibinfo
  {volume} {109}},\ \bibinfo {pages} {9770--9774} (\bibinfo {year}
  {2012})}\BibitemShut {NoStop}%
\bibitem [{\citenamefont {Robens}\ \emph {et~al.}(2017)\citenamefont {Robens},
  \citenamefont {Zopes}, \citenamefont {Alt}, \citenamefont {Brakhane},
  \citenamefont {Meschede},\ and\ \citenamefont {Alberti}}]{Robens2017}%
  \BibitemOpen
  \bibfield  {author} {\bibinfo {author} {\bibfnamefont {C.}~\bibnamefont
  {Robens}}, \bibinfo {author} {\bibfnamefont {J.}~\bibnamefont {Zopes}},
  \bibinfo {author} {\bibfnamefont {W.}~\bibnamefont {Alt}}, \bibinfo {author}
  {\bibfnamefont {S.}~\bibnamefont {Brakhane}}, \bibinfo {author}
  {\bibfnamefont {D.}~\bibnamefont {Meschede}}, \ and\ \bibinfo {author}
  {\bibfnamefont {A.}~\bibnamefont {Alberti}},\ }\bibfield  {title} {\enquote
  {\bibinfo {title} {Low-entropy states of neutral atoms in
  polarization-synthesized optical lattices},}\ }\href {\doibase
  10.1103/PhysRevLett.118.065302} {\bibfield  {journal} {\bibinfo  {journal}
  {Phys. Rev. Lett.}\ }\textbf {\bibinfo {volume} {118}},\ \bibinfo {pages}
  {065302} (\bibinfo {year} {2017})}\BibitemShut {NoStop}%
\bibitem [{\citenamefont {Robens}\ \emph {et~al.}(2018)\citenamefont {Robens},
  \citenamefont {Brakhane}, \citenamefont {Alt}, \citenamefont {Meschede},
  \citenamefont {Zopes},\ and\ \citenamefont {Alberti}}]{Robens2018}%
  \BibitemOpen
  \bibfield  {author} {\bibinfo {author} {\bibfnamefont {Carsten}\ \bibnamefont
  {Robens}}, \bibinfo {author} {\bibfnamefont {Stefan}\ \bibnamefont
  {Brakhane}}, \bibinfo {author} {\bibfnamefont {Wolfgang}\ \bibnamefont
  {Alt}}, \bibinfo {author} {\bibfnamefont {Dieter}\ \bibnamefont {Meschede}},
  \bibinfo {author} {\bibfnamefont {Jonathan}\ \bibnamefont {Zopes}}, \ and\
  \bibinfo {author} {\bibfnamefont {Andrea}\ \bibnamefont {Alberti}},\
  }\bibfield  {title} {\enquote {\bibinfo {title} {Fast, high-precision optical
  polarization synthesizer for ultracold-atom experiments},}\ }\href {\doibase
  10.1103/PhysRevApplied.9.034016} {\bibfield  {journal} {\bibinfo  {journal}
  {Phys. Rev. Applied}\ }\textbf {\bibinfo {volume} {9}},\ \bibinfo {pages}
  {034016} (\bibinfo {year} {2018})}\BibitemShut {NoStop}%
\bibitem [{\citenamefont {Gl\"uck}\ \emph {et~al.}(2002)\citenamefont
  {Gl\"uck}, \citenamefont {Kolovsky},\ and\ \citenamefont
  {Korsch}}]{GLUCK2002}%
  \BibitemOpen
  \bibfield  {author} {\bibinfo {author} {\bibfnamefont {M.}~\bibnamefont
  {Gl\"uck}}, \bibinfo {author} {\bibfnamefont {A.~R.}\ \bibnamefont
  {Kolovsky}}, \ and\ \bibinfo {author} {\bibfnamefont {H.~J.}\ \bibnamefont
  {Korsch}},\ }\bibfield  {title} {\enquote {\bibinfo {title} {Wannier-stark
  resonances in optical and semiconductor superlattices},}\ }\href {\doibase
  https://doi.org/10.1016/S0370-1573(02)00142-4} {\bibfield  {journal}
  {\bibinfo  {journal} {Phys. Rep.}\ }\textbf {\bibinfo {volume} {366}},\
  \bibinfo {pages} {103 -- 182} (\bibinfo {year} {2002})}\BibitemShut {NoStop}%
\bibitem [{\citenamefont {Kitagawa}\ and\ \citenamefont
  {Ueda}(1993)}]{Kitagawa1993}%
  \BibitemOpen
  \bibfield  {author} {\bibinfo {author} {\bibfnamefont {M.}~\bibnamefont
  {Kitagawa}}\ and\ \bibinfo {author} {\bibfnamefont {M.}~\bibnamefont
  {Ueda}},\ }\bibfield  {title} {\enquote {\bibinfo {title} {Squeezed spin
  states},}\ }\href {\doibase 10.1103/PhysRevA.47.5138} {\bibfield  {journal}
  {\bibinfo  {journal} {Phys. Rev. A}\ }\textbf {\bibinfo {volume} {47}},\
  \bibinfo {pages} {5138--5143} (\bibinfo {year} {1993})}\BibitemShut {NoStop}%
\bibitem [{\citenamefont {Gross}(2012)}]{Gross2012a}%
  \BibitemOpen
  \bibfield  {author} {\bibinfo {author} {\bibfnamefont {C.}~\bibnamefont
  {Gross}},\ }\bibfield  {title} {\enquote {\bibinfo {title} {Spin squeezing,
  entanglement and quantum metrology with bose-einstein condensates},}\ }\href
  {http://stacks.iop.org/0953-4075/45/i=10/a=103001} {\bibfield  {journal}
  {\bibinfo  {journal} {J. Phys. B At. Mol. Opt. Phys.}\ }\textbf {\bibinfo
  {volume} {45}},\ \bibinfo {pages} {103001} (\bibinfo {year}
  {2012})}\BibitemShut {NoStop}%
\bibitem [{\citenamefont {Hosten}\ \emph {et~al.}(2016)\citenamefont {Hosten},
  \citenamefont {Engelsen}, \citenamefont {Krishnakumar},\ and\ \citenamefont
  {Kasevich}}]{Hosten2016}%
  \BibitemOpen
  \bibfield  {author} {\bibinfo {author} {\bibfnamefont {O.}~\bibnamefont
  {Hosten}}, \bibinfo {author} {\bibfnamefont {N.~J.}\ \bibnamefont
  {Engelsen}}, \bibinfo {author} {\bibfnamefont {R.}~\bibnamefont
  {Krishnakumar}}, \ and\ \bibinfo {author} {\bibfnamefont {M.~A.}\
  \bibnamefont {Kasevich}},\ }\bibfield  {title} {\enquote {\bibinfo {title}
  {Measurement noise 100 times lower than the quantum-projection limit using
  entangled atoms},}\ }\href {\doibase 10.1038/nature16176} {\bibfield
  {journal} {\bibinfo  {journal} {Nature}\ }\textbf {\bibinfo {volume} {529}},\
  \bibinfo {pages} {505--508} (\bibinfo {year} {2016})}\BibitemShut {NoStop}%
\bibitem [{\citenamefont {Wineland}\ \emph {et~al.}(1992)\citenamefont
  {Wineland}, \citenamefont {Bollinger}, \citenamefont {Itano}, \citenamefont
  {Moore},\ and\ \citenamefont {Heinzen}}]{Wineland1992}%
  \BibitemOpen
  \bibfield  {author} {\bibinfo {author} {\bibfnamefont {D.~J.}\ \bibnamefont
  {Wineland}}, \bibinfo {author} {\bibfnamefont {J.~J.}\ \bibnamefont
  {Bollinger}}, \bibinfo {author} {\bibfnamefont {W.~M.}\ \bibnamefont
  {Itano}}, \bibinfo {author} {\bibfnamefont {F.~L.}\ \bibnamefont {Moore}}, \
  and\ \bibinfo {author} {\bibfnamefont {D.~J.}\ \bibnamefont {Heinzen}},\
  }\bibfield  {title} {\enquote {\bibinfo {title} {Spin squeezing and reduced
  quantum noise in spectroscopy},}\ }\href {\doibase 10.1103/PhysRevA.46.R6797}
  {\bibfield  {journal} {\bibinfo  {journal} {Phys. Rev. A}\ }\textbf {\bibinfo
  {volume} {46}},\ \bibinfo {pages} {R6797--R6800} (\bibinfo {year}
  {1992})}\BibitemShut {NoStop}%
\bibitem [{\citenamefont {Sorensen}\ \emph {et~al.}(2001)\citenamefont
  {Sorensen}, \citenamefont {Duan}, \citenamefont {Cirac},\ and\ \citenamefont
  {Zoller}}]{Sorensen2001}%
  \BibitemOpen
  \bibfield  {author} {\bibinfo {author} {\bibfnamefont {A.}~\bibnamefont
  {Sorensen}}, \bibinfo {author} {\bibfnamefont {L.~M.}\ \bibnamefont {Duan}},
  \bibinfo {author} {\bibfnamefont {J.~I.}\ \bibnamefont {Cirac}}, \ and\
  \bibinfo {author} {\bibfnamefont {P.}~\bibnamefont {Zoller}},\ }\bibfield
  {title} {\enquote {\bibinfo {title} {Many-particle entanglement with
  bose-einstein condensates},}\ }\href {\doibase 10.1038/35051038} {\bibfield
  {journal} {\bibinfo  {journal} {Nature}\ }\textbf {\bibinfo {volume} {409}},\
  \bibinfo {pages} {63--66} (\bibinfo {year} {2001})}\BibitemShut {NoStop}%
\bibitem [{\citenamefont {Ma}\ \emph {et~al.}(2011)\citenamefont {Ma},
  \citenamefont {Wang}, \citenamefont {Sun},\ and\ \citenamefont
  {Nori}}]{MA2011}%
  \BibitemOpen
  \bibfield  {author} {\bibinfo {author} {\bibfnamefont {J.}~\bibnamefont
  {Ma}}, \bibinfo {author} {\bibfnamefont {X.}~\bibnamefont {Wang}}, \bibinfo
  {author} {\bibfnamefont {C.P.}\ \bibnamefont {Sun}}, \ and\ \bibinfo {author}
  {\bibfnamefont {F.}~\bibnamefont {Nori}},\ }\bibfield  {title} {\enquote
  {\bibinfo {title} {Quantum spin squeezing},}\ }\href {\doibase
  http://dx.doi.org/10.1016/j.physrep.2011.08.003} {\bibfield  {journal}
  {\bibinfo  {journal} {Phys. Rep.}\ }\textbf {\bibinfo {volume} {509}},\
  \bibinfo {pages} {89 -- 165} (\bibinfo {year} {2011})}\BibitemShut {NoStop}%
\bibitem [{\citenamefont {Gross}\ \emph {et~al.}(2010)\citenamefont {Gross},
  \citenamefont {Zibold}, \citenamefont {Nicklas}, \citenamefont {Est\'eve},\
  and\ \citenamefont {Oberthaler}}]{Gross2010}%
  \BibitemOpen
  \bibfield  {author} {\bibinfo {author} {\bibfnamefont {C.}~\bibnamefont
  {Gross}}, \bibinfo {author} {\bibfnamefont {T.}~\bibnamefont {Zibold}},
  \bibinfo {author} {\bibfnamefont {E.}~\bibnamefont {Nicklas}}, \bibinfo
  {author} {\bibfnamefont {J.}~\bibnamefont {Est\'eve}}, \ and\ \bibinfo
  {author} {\bibfnamefont {M.~K.}\ \bibnamefont {Oberthaler}},\ }\bibfield
  {title} {\enquote {\bibinfo {title} {Nonlinear atom interferometer surpasses
  classical precision limit},}\ }\href {\doibase 10.1038/nature08919}
  {\bibfield  {journal} {\bibinfo  {journal} {Nature}\ }\textbf {\bibinfo
  {volume} {464}},\ \bibinfo {pages} {1165--1169} (\bibinfo {year}
  {2010})}\BibitemShut {NoStop}%
\bibitem [{\citenamefont {Riedel}\ \emph {et~al.}(2010)\citenamefont {Riedel},
  \citenamefont {B\"ohi}, \citenamefont {Li}, \citenamefont {H\"ansch},
  \citenamefont {Sinatra},\ and\ \citenamefont {Treutlein}}]{Riedel2010}%
  \BibitemOpen
  \bibfield  {author} {\bibinfo {author} {\bibfnamefont {M.~F.}\ \bibnamefont
  {Riedel}}, \bibinfo {author} {\bibfnamefont {P.}~\bibnamefont {B\"ohi}},
  \bibinfo {author} {\bibfnamefont {Y.}~\bibnamefont {Li}}, \bibinfo {author}
  {\bibfnamefont {Theodor~W.}\ \bibnamefont {H\"ansch}}, \bibinfo {author}
  {\bibfnamefont {A.}~\bibnamefont {Sinatra}}, \ and\ \bibinfo {author}
  {\bibfnamefont {P.}~\bibnamefont {Treutlein}},\ }\bibfield  {title} {\enquote
  {\bibinfo {title} {Atom-chip-based generation of entanglement for quantum
  metrology},}\ }\href {\doibase 10.1038/nature08988} {\bibfield  {journal}
  {\bibinfo  {journal} {Nature}\ }\textbf {\bibinfo {volume} {464}},\ \bibinfo
  {pages} {1170--1173} (\bibinfo {year} {2010})}\BibitemShut {NoStop}%
\bibitem [{\citenamefont {Dimopoulos}\ \emph {et~al.}(2009)\citenamefont
  {Dimopoulos}, \citenamefont {Graham}, \citenamefont {Hogan}, \citenamefont
  {Kasevich},\ and\ \citenamefont {Rajendran}}]{Dimopoulos}%
  \BibitemOpen
  \bibfield  {author} {\bibinfo {author} {\bibfnamefont {S.}~\bibnamefont
  {Dimopoulos}}, \bibinfo {author} {\bibfnamefont {P.~W.}\ \bibnamefont
  {Graham}}, \bibinfo {author} {\bibfnamefont {J.~M.}\ \bibnamefont {Hogan}},
  \bibinfo {author} {\bibfnamefont {M.~A.}\ \bibnamefont {Kasevich}}, \ and\
  \bibinfo {author} {\bibfnamefont {S.}~\bibnamefont {Rajendran}},\ }\bibfield
  {title} {\enquote {\bibinfo {title} {Gravitational wave detection with atom
  interferometry},}\ }\href {\doibase
  https://doi.org/10.1016/j.physletb.2009.06.011} {\bibfield  {journal}
  {\bibinfo  {journal} {Phys. Lett. B}\ }\textbf {\bibinfo {volume} {678}},\
  \bibinfo {pages} {37 -- 40} (\bibinfo {year} {2009})}\BibitemShut {NoStop}%
\bibitem [{\citenamefont {Sukenik}\ \emph {et~al.}(1993)\citenamefont
  {Sukenik}, \citenamefont {Boshier}, \citenamefont {Cho}, \citenamefont
  {Sandoghdar},\ and\ \citenamefont {Hinds}}]{Sukenik}%
  \BibitemOpen
  \bibfield  {author} {\bibinfo {author} {\bibfnamefont {C.~I.}\ \bibnamefont
  {Sukenik}}, \bibinfo {author} {\bibfnamefont {M.~G.}\ \bibnamefont
  {Boshier}}, \bibinfo {author} {\bibfnamefont {D.}~\bibnamefont {Cho}},
  \bibinfo {author} {\bibfnamefont {V.}~\bibnamefont {Sandoghdar}}, \ and\
  \bibinfo {author} {\bibfnamefont {E.~A.}\ \bibnamefont {Hinds}},\ }\bibfield
  {title} {\enquote {\bibinfo {title} {Measurement of the casimir-polder
  force},}\ }\href {\doibase 10.1103/PhysRevLett.70.560} {\bibfield  {journal}
  {\bibinfo  {journal} {Phys. Rev. Lett.}\ }\textbf {\bibinfo {volume} {70}},\
  \bibinfo {pages} {560--563} (\bibinfo {year} {1993})}\BibitemShut {NoStop}%
\bibitem [{\citenamefont {Haslinger}\ \emph {et~al.}(2018)\citenamefont
  {Haslinger}, \citenamefont {Jaffe}, \citenamefont {Xu}, \citenamefont
  {Schwartz}, \citenamefont {Sonnleitner}, \citenamefont {Ritsch-Marte},
  \citenamefont {Ritsch},\ and\ \citenamefont {M{\"u}ller}}]{Haslinger}%
  \BibitemOpen
  \bibfield  {author} {\bibinfo {author} {\bibfnamefont {P.}~\bibnamefont
  {Haslinger}}, \bibinfo {author} {\bibfnamefont {M.}~\bibnamefont {Jaffe}},
  \bibinfo {author} {\bibfnamefont {V.}~\bibnamefont {Xu}}, \bibinfo {author}
  {\bibfnamefont {O.}~\bibnamefont {Schwartz}}, \bibinfo {author}
  {\bibfnamefont {M.}~\bibnamefont {Sonnleitner}}, \bibinfo {author}
  {\bibfnamefont {M.}~\bibnamefont {Ritsch-Marte}}, \bibinfo {author}
  {\bibfnamefont {H.}~\bibnamefont {Ritsch}}, \ and\ \bibinfo {author}
  {\bibfnamefont {H.}~\bibnamefont {M{\"u}ller}},\ }\bibfield  {title}
  {\enquote {\bibinfo {title} {Attractive force on atoms due to blackbody
  radiation},}\ }\href {\doibase 10.1038/s41567-017-0004-9} {\bibfield
  {journal} {\bibinfo  {journal} {Nat. Phys.}\ }\textbf {\bibinfo {volume}
  {14}},\ \bibinfo {pages} {257--260} (\bibinfo {year} {2018})}\BibitemShut
  {NoStop}%
\end{thebibliography}%

\end{document}